\documentclass[12pt,twoside, onecolumn,draftclsnofoot]{IEEEtran}
\newtheorem{proposition}{{Proposition}}

\newtheorem{remark}{{Remark}}

\usepackage{cite}
\usepackage{amsmath,amssymb,amsfonts}
\usepackage{algorithmic}
\usepackage{graphicx}
\usepackage{textcomp}
\usepackage{xcolor}
\usepackage{ulem}

\normalsize

\ifCLASSINFOpdf
\else
\fi

\begin{document}

\title{Constructive Approaches to Perception-Aware Lossy Source Coding: Information-Theoretic Guidelines}

\author{Ali~Hussein, Jun~Chen, Chao~Tian, S.~Sandeep~Pradhan\thanks{Ali Hussein and Jun Chen are with the Department  of Electrical and Computer Engineering, McMaster University, Hamilton, ON L8S 4K1, Canada (e-mail: hussea34@mcmaster.ca; chenjun@mcmaster.ca).}\thanks{Chao Tian is with the Department of Electrical and Computer Engineering, Texas A\&M University,  College Station, TX 77843-3128, USA (e-mail:
		chao.tian@tamu.edu).}\thanks{S. Sandeep Pradhan is with the Department of Electrical Engineering and Computer Science, University of Michigan, Ann Arbor, MI 48109, USA (e-mail: pradhanv@umich.edu).}}

\maketitle

\begin{abstract}
	%\boldmath
	
Perception-aware lossy source coding has attracted significant recent interest. It augments the classical distortion criterion with an explicit perception constraint, thereby enabling more refined control over fidelity and perceptual quality. Despite rapid progress, the diversity of rate–distortion–perception formulations and their underlying assumptions remains poorly understood by many practitioners. In particular, there is often a tendency to rely heavily on the expressive power of deep neural networks and generative models without clear theoretical guidance, using fundamental limits merely as performance benchmarks rather than as sources of design insight.

This tutorial paper aims to bridge this gap by surveying information-theoretic principles that can be leveraged to develop constructive approaches to perception-aware lossy source coding. We distill practical guidelines implied by rate–distortion–perception theory and demonstrate how they inform the design of implementable coding schemes. A simple unit-circle example is used as a pedagogical tool to  illustrate key ideas, architectural principles, and tradeoffs in an intuitive and unified manner. Both one-shot and asymptotic  settings are examined to highlight conceptual similarities and operational differences.
We also clarify the role of common randomness and the notion of universal representation, and elucidate the connections between perception-aware and conventional lossy source coding. Overall, this tutorial provides a principled foundation for developing perception-aware compression systems that go beyond black-box model design.

	%have corrected: outperfrom -> outperform need to update the abstract when submitting the revised version
	
	% when specialized to the scenario where %common randomness is absent and 
	%the perception constraint is inactive.
\end{abstract}


\IEEEpeerreviewmaketitle

\section{Introduction}

It has long been recognized \cite{BM18} that distortion alone does not fully capture human perception of the quality of a compressed signal. Against this backdrop, perception-aware lossy source coding augments the classical distortion criterion with an explicit perception constraint, thereby enabling more refined control over fidelity and perceptual quality. This additional constraint reflects the requirement that reconstructed signals should not only be close to the source under a prescribed distortion measure, but also appear natural or realistic according to the statistics of the source. As a result, the design of compression systems must account for both signal fidelity and distributional consistency.

On the theoretical side, incorporating a perception (also referred to as realism) constraint calls for an extension of Shannon’s rate-distortion theory \cite{CT91}. In practice, perception loss is often quantified via a divergence between the source and reconstruction distributions. While related ideas can be traced back to earlier work on distribution-preserving quantization \cite{LKK10, KZLK13} and output-constrained lossy source coding \cite{SLY15J1, SLY15J2}, the recent surge of interest in rate–distortion–perception theory can be largely attributed to the thought-provoking paper by Blau and Michaeli \cite{BM19}. Their work opened the floodgates for a systematic investigation of the fundamental tradeoffs among rate, distortion, and perceptual quality, spurring a rapidly growing body of research in this area \cite{Matsumoto18,Matsumoto19,YWYML21,TW21,TA21,ZQCK21,QZCK22,LZCK22,YWL22,LZCK22J,CYWSGT22,Wagner22,YG23,NGBH23,SPCYK23,SSK23,HWG24,ZT24,GPC24,SCKY24,XLCZ24,XLCYZ25,QSCKYSGT24,QCYX25,XLCYZ24}. By formalizing the role of perception within an information-theoretic framework, it has helped shape a new direction for the field, marking the beginning of a distinct and influential line of inquiry in modern information theory \cite{CFKOS25}.

However, despite rapid theoretical progress, its impact on practice remains limited. In particular, practitioners of learned image compression tend to rely almost exclusively on the expressive power of deep neural networks and generative models, often without seeking guidance from information-theoretic principles. This disconnect can be attributed to two main factors. First, information-theoretic limits are frequently used merely as performance benchmarks, with their rich operational implications largely overlooked. Second, recent studies \cite{CYWSGT22,SCKY24,XLCZ24,XLCYZ24} have shown that the fundamental rate–distortion–perception tradeoffs depend critically on the underlying problem formulation, rendering many theoretical results difficult to interpret without a solid background in information theory.

This tutorial paper aims to bridge this gap by presenting  the information-theoretic guidelines for perception-aware lossy source coding revealed by recent advances in rate–distortion–perception theory. Our focus is on the squared-error distortion measure and the squared Wasserstein-$2$ perception measure, for which the theory is currently the most well developed. A simple unit-circle example is employed as a pedagogical tool, through which the essential constructions and design principles can be introduced in a transparent and intuitive manner. Building on the insights gained from this example, we then extend the discussion to both the one-shot and asymptotic settings for general source distributions, highlighting their conceptual similarities as well as their operational differences. The role of common randomness, a subtle yet crucial component that is often a source of confusion in both theory and practice, is  clarified. We also elucidate the notion of universal representation, which is firmly grounded in rate–distortion–perception theory and carries clear operational significance, enabling traversal of the entire optimal distortion–perception tradeoff curve with a fixed encoder.
It is our hope that this tutorial will make these information-theoretic insights more accessible to practitioners and foster a more principled approach to the design of perception-aware lossy source coding systems.

The remainder of this paper is organized as follows. Section \ref{sec:example} introduces a simple unit-circle example, systematically illustrating the design procedures for handling increasingly complex formulations;  starting from the case of a perfect perception constraint and no common randomness, we gradually consider imperfect perception constraints and limited common randomness to explain how each added complexity affects the design. Section \ref{sec:one-shot} addresses the one-shot setting, in which the system operates over a single source variable, potentially high-dimensional, and demonstrates how neural compression and rectified flow can be leveraged for the compressive and generative tasks in perception-aware lossy source coding. Section \ref{sec:asymptotic} focuses on the asymptotic setting, where long sequences of i.i.d. source variables are encoded and reconstructed subject to both distortion and perception constraints; the associated constructions are presented using both random  and structured codes. Finally, Section \ref{sec:conclusion} concludes the paper with some discussions.

We adopt the following notation throughout the paper. Uppercase letters typically denote random variables or vectors, while their lowercase counterparts represent realizations. The distribution of a random variable or vector 
$X$ is denoted by 
$p_X$. For two distributions 
$p_X$ and 
$p_Y$ over 
$\mathbb{R}^k$, the Wasserstein-$2$
 distance is defined as
\begin{align}
	W_2(p_X,p_{Y}):=\inf\limits_{p_{XY}\in\Pi(p_X,p_Y)}(\mathbb{E}[\|X-Y\|^2])^{\frac{1}{2}},
\end{align}
where $\Pi(p_X,p_Y)$ denotes the set of all joint distributions with marginals $p_X$ and $p_Y$. We write 
$\mathbb{E}[\cdot]$, $H(\cdot)$, and $I(\cdot;\cdot)$ for expectation, entropy, and mutual information, respectively; when necessary, subscripts are added to indicate the underlying distribution. The notation 
$(a)_+$ stands for 
$\max\{a,0\}$. All logarithms are taken to base 
$2$.
 
%We typically use uppercase letters for random variables/vectors and lowercase counterparts for their realizations. The distribution of a random variable/vector $X$ is written as $p_X$. 
%For two distributions $p_X$ and $p_Y$ over $\mathbb{R}^k$, the
%Wasserstein-$2$ distance $W_2(p_X,p_{\hat{X}})$ is defined as

%where $\Pi(p_X,p_Y)$ denotes the set of joint distributions with marginals $p_X$ and $p_Y$. We use $\mathbb{E}[\cdot]$, $H(\cdot)$, and $I(\cdot;\cdot)$ to denote expectation, entropy, and mutual information, respective; sometimes 
%we add subscript to them to highlight the underlying distribution. 
%$(a)_+$ denotes $\max\{a,0\}$.
%Throughout this paper, the logarithm function is to base $2$.

\section{A Unit-Circle Example}\label{sec:example}

Consider a toy example in which the source variable $X$ is uniformly distributed over the unit circle $\mathbb{S}:=\{s\in\mathbb{R}^2:\|s\|=1\}$. Let $\hat{X}$ denote the reconstructed variable. The distortion and perception losses are quantified by the mean squared error $\mathbb{E}[\|X-\hat{X}\|^2]$ and the squared Wasserstein-$2$ distance $W^2_2(p_X,p_{\hat{X}})$, respectively. A preliminary version of this example was studied in \cite{TA21}; see also \cite{ZT24} for a variant.

\subsection{$1$-bit Quantization}

\begin{figure}[htbp]
	\centerline{\includegraphics[width=8.5cm]{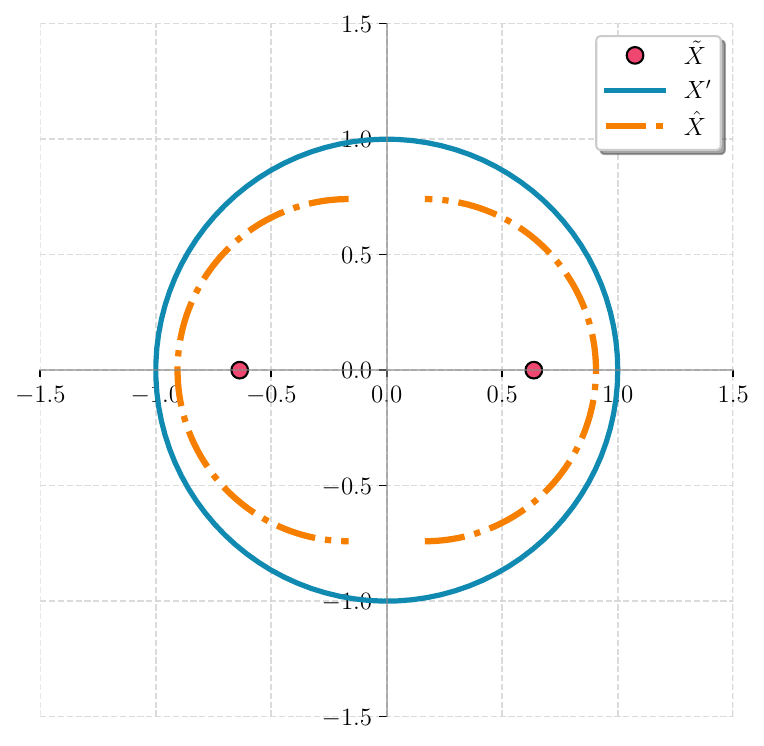}} \caption{Illustration of the supports of $\tilde{X}$, $X'$, and $\hat{X}$ (with $P=0.04$) in the unit-circle example for $1$-bit quantization. Specifically, $\tilde{X}$ is uniformly distributed over the two points $(\pm\frac{2}{\pi},0)\approx(\pm0.637,0)$, $X'$ is uniformly distributed over the unit circle centered at the origin, and $\hat{X}$ is uniformly distributed over two semicircles of radius $1-\frac{\pi}{5\sqrt{\pi^2-4}}\approx 0.741$,  centered at $(\pm\frac{2}{5\sqrt{\pi^2-4}},0)\approx(\pm 0.165,0)$, respectively.}\label{fig:2*1}
\end{figure}

Now suppose we perform a $1$-bit quantization of $X$. Specifically, let $J:=f(X)$, where $f:\mathbb{S}\rightarrow\{0,1\}$ is a (possibly stochastic) encoder, and let  $\tilde{X}:=\mathbb{E}[X|J]$ denote the MMSE estimate of $X$ given $J$. By symmetry, an optimal encoder (see \cite[Appendix C]{TA21} for a rigorous proof) that minimizes $\mathbb{E}[\|X-\tilde{X}\|^2]$ partitions the unit circle into two halves, mapping the right unit semicircle to $0$ and the left unit semicircle to $1$.
In this case, $\tilde{X}$ equals the centroid of the corresponding semicircle, taking the value $(\frac{2}{\pi},0)$ when $J=0$ and $(-\frac{2}{\pi},0)$ when $J=1$ (see Fig. \ref{fig:2*1}).
A direct calculation shows that
\begin{align}
	\mathbb{E}[\|X-\tilde{X}\|^2]=1-\frac{4}{\pi^2}.\label{eq:distortion_1bit}
\end{align}
Interestingly, the greedy transport plan that maps each point on the unit circle to its nearest centroid---namely, mapping the right unit semicircle to $(\frac{2}{\pi},0)$ and the left unit semicircle to $(-\frac{2}{\pi},0)$---induces exactly the distribution $p_{\tilde{X}}$. Consequently,
\begin{align}
W^2_2(p_X,p_{\tilde{X}})=1-\frac{4}{\pi^2},\label{eq:perception_1bit}
\end{align}
implying that, when 
$\tilde{X}$
is used as the reconstruction, the distortion and perception losses are equal. This is not a coincidence but rather an instance of a more general phenomenon, as formalized in the following proposition.

\begin{proposition}\label{prop:1}
	Let  $Z$ be a random vector in $\mathbb{R}^k$ with $\mathbb{E}[\|Z\|^2]<\infty$.
	Fix a positive integer $M$, and consider the class of all (possibly stochastic) mappings
	$\phi:\mathbb{R}^k\rightarrow\{0,1,\ldots,M-1\}$. For each such mapping, define the MMSE estimate
	 $\tilde{Z}:=\mathbb{E}[Z|\phi(Z)]$.
	 Let $\phi^*$ be a minimizer of 
	 $\mathbb{E}[\|Z-\tilde{Z}\|^2]$ over this class, and denote the corresponding estimate 
	  by $\tilde{Z}^*$.
	Then,
	\begin{align}
		\mathbb{E}[\|Z-\tilde{Z}^*\|^2]=W^2_2(p_Z,p_{\tilde{Z}^*}).
	\end{align}
\end{proposition}
\begin{remark}
Proposition \ref{prop:1} admits a simple proof. Indeed, if $\mathbb{E}[\|Z-\tilde{Z}^*\|^2]>W^2_2(p_Z,p_{\tilde{Z}^*})$, then the coupling achieving $W^2_2(p_Z,p_{\tilde{Z}^*})$ induces a (possibly stochastic) mapping that yields a strictly smaller distortion than $\phi^*$, thereby contradicting the optimality of $\phi^*$.
\end{remark}

In the $1$-bit quantization scenario, a simple way to generate a perceptually perfect sample 
$X'$ 
satisfying $W^2_2(p_X,p_{X'})=0$
 (equivalently, 
$p_{X'}=p_X$), is to let the decoder 
$g$ act as a stochastic inverse of the encoder $f$, i.e., to perform posterior sampling. Specifically, given 
$J$, the reconstruction 
$X':=g(J)$ is drawn uniformly from the corresponding semicircle: the right semicircle when 
$J=0$ and the left semicircle when 
$J=1$. Such a decoder ensures that the reconstruction distribution matches the source distribution, thereby achieving perfect perceptual quality. However, this comes at the expense of increased distortion: the resulting mean squared error is exactly twice that of the MMSE estimate 
$\tilde{X}$, i.e.,
\begin{align}
	\mathbb{E}[\|X-X'\|^2]=2\mathbb{E}[\|X-\tilde{X}\|^2].
\end{align}
It turns out that this is the minimum achievable distortion subject to the perfect perception constraint. Moreover, this relationship is not coincidental, but rather reflects a more general phenomenon, as formalized in the following proposition \cite[Theorem 2]{YWYML21}\cite[Theorem 2]{TA21}.

\begin{proposition}\label{prop:2}
Let  $Z$ be a random vector in $\mathbb{R}^k$ with $\mathbb{E}[\|Z\|^2]<\infty$.
Fix a positive integer $M$, and consider the class of all (possibly stochastic) encoder–decoder pairs
$\phi:\mathbb{R}^k\rightarrow\{0,1,\ldots,M-1\}$ and $\psi:\{0,1,\ldots,M-1\}\rightarrow\mathbb{R}^k$. Define $Z':=\psi(\phi(Z))$ and $\tilde{Z}:=\mathbb{E}[X|\phi(Z)]$. Then,
\begin{align}
	\min\limits_{\phi,\psi: p_{Z'}=p_Z}\mathbb{E}[\|Z-Z'\|^2]=2\min\limits_{\phi}\mathbb{E}[\|Z-\tilde{Z}\|^2].
\end{align}
\end{proposition}
\begin{remark}
	Proposition \ref{prop:2} also admits a simple proof. By the orthogonality property of the MMSE estimate and the constraint $p_{Z'}=p_Z$, 
	\begin{align}
		\mathbb{E}[\|Z-Z'\|^2]&=\mathbb{E}[\|Z-\tilde{Z}\|^2]+\mathbb{E}[\|Z'-\tilde{Z}\|^2]\nonumber\\
		&\geq 2\min\{\mathbb{E}[\|Z-\tilde{Z}\|^2],\mathbb{E}[\|Z'-\tilde{Z}\|^2]\}\nonumber\\
		&\geq 2\min\limits_{\phi}\mathbb{E}[\|Z-\tilde{Z}\|^2].
	\end{align}
To attain this lower bound, the encoder should be chosen to minimize  $\mathbb{E}[\|Z-\tilde{Z}\|]^2$, while the decoder is constructed as a stochastic inverse of the encoder.
\end{remark}

We now turn to the case of an arbitrary perception constraint $W^2_2(p_X,p_{\hat{X}})\leq P$.
A simple strategy is to construct $\hat{X}$ by linearly interpolating between  the MMSE estimate $\tilde{X}$ and the perceptually perfect sample $X'$, i.e.,
\begin{align}
	\hat{X}:=(1-\alpha)X'+\alpha\tilde{X}, 
\end{align}
where the parameter $\alpha\in[0,1]$ is chosen such that $W^2_2(p_X,p_{\hat{X}})=P$.
Remarkably, despite its apparent simplicity, this approach is in fact optimal and achieves the best possible distortion–perception tradeoff, as formalized in the following proposition \cite[Theorems 1 and 3]{FMM21}.

\begin{figure}[htbp]
	\centerline{\includegraphics[width=12cm]{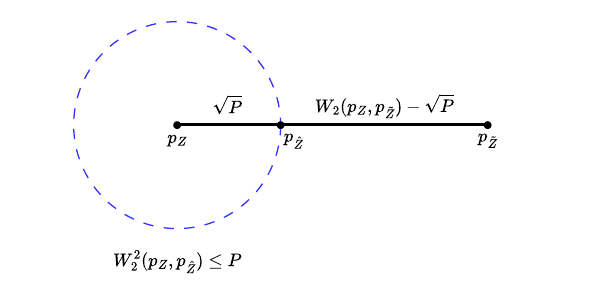}} \caption{Due to the perception constraint, $p_{\hat{X}}$ must lie within the Wasserstein-$2$ ball of radius $\sqrt{P}$ centered at $p_Z$. Among all such $p_{\hat{Z}}$, the one that minimizes $W _2(p_{\hat{Z}},p_{\tilde{Z}})$ is located at the point where the geodesic from $p_Z$ to $p_{\tilde{Z}}$ intersects the boundary of this ball. For this choice of $p_{\hat{Z}}$, we have $W_2(p_{\hat{Z}},p_{\tilde{Z}})=W_2(p_Z,p_{\tilde{Z}})-\sqrt{P}$. }\label{fig:interpolation}
\end{figure}

\begin{proposition}\label{prop:interpolation}
Consider a random vector $Z$ in $\mathbb{R}^k$ with $\mathbb{E}[\|Z\|^2]<\infty$, and a representation $W$ generated according to some conditional distribution $p_{W|Z}$, and a reconstruction $\hat{Z}$  based on $W$. Suppose that the perception constraint $W^2_2(p_Z,p_{\hat{Z}})\leq P$ is satisfied. Then,
\begin{align}
	\mathbb{E}[\|Z-\hat{Z}\|^2]\geq\mathbb{E}[\|Z-\tilde{Z}\|^2]+[(W_2(p_Z,p_{\tilde{Z}})-\sqrt{P})_+]^2,\label{eq:DPlowerbound}
\end{align}
with $\tilde{Z}:=\mathbb{E}[Z|W]$. Moreover, the lower bound in \eqref{eq:DPlowerbound} is attained by the following interpolation scheme: %\footnote{Here, we assume $W_2(p_Z,p_{\tilde{Z}})>0$; otherwise, $\tilde{Z}=Z$ a.s.}
\begin{align}
	\hat{Z}:=\begin{cases}
		\left(1-\frac{\sqrt{P}}{W_2(p_Z,p_{\tilde{Z}})}\right)Z'+\frac{\sqrt{P}}{W_2(p_Z,p_{\tilde{Z}})}\tilde{Z}, &P\in[0,W^2_2(p_Z,p_{\tilde{Z}})),\\
		\tilde{Z},& P\geq W^2_2(p_Z,p_{\tilde{Z}}),
	\end{cases}\label{eq:optimaldecoder}
\end{align}
where $Z'$ is jointly distributed with $(Z,W)$ such that $Z\leftrightarrow W\leftrightarrow Z'$ forms a Markov chain, $p_{Z'}=p_Z$, and $\mathbb{E}[\|Z'-\tilde{Z}\|^2]=W^2_2(p_Z,p_{\tilde{Z}})$. 
\end{proposition}
\begin{remark}
	Proposition \ref{prop:interpolation} admits a natural geometric interpretation \cite[Theorem 4]{ZQCK21}\cite[Figure 2]{FMM21} (see Fig. \ref{fig:interpolation} for an illustration). By the orthogonality property of the MMSE estimate,
	\begin{align}
		\mathbb{E}[\|Z-\hat{Z}\|^2]&=\mathbb{E}[\|Z-\tilde{Z}\|^2]+\mathbb{E}[\|\hat{Z}-\tilde{Z}\|^2]\nonumber\\
		&\geq\mathbb{E}[\|Z-\tilde{Z}\|^2]+W^2_2(p_{\hat{Z}},p_{\tilde{Z}}).\label{eq:orthogonal}
	\end{align}
	Due to the perception constraint, $p_{\hat{Z}}$ lies in the Wasserstein-$2$ ball of radius $\sqrt{P}$ centered at $p_{Z}$, i.e., $W_2(p_Z,p_{\hat{Z}})\leq\sqrt{P}$. By the triangle inequality for the Wasserstein-$2$ distance, 
	\begin{align}
		W_2(p_{\hat{Z}},p_{\tilde{Z}})&\geq W_2(p_Z,p_{\tilde{Z}})-W_2(p_Z,p_{\hat{Z}})\nonumber\\
		&\geq W_2(p_Z,p_{\tilde{Z}})-\sqrt{P}.\label{eq:triangle}
	\end{align}
	Substituting \eqref{eq:triangle} into \eqref{eq:orthogonal} and invoking the nonnegativity of the Wasserstein-$2$ distance yields \eqref{eq:DPlowerbound}. It remains to verify that the reconstruction $\hat{Z}$ induced by the interpolation scheme in \eqref{eq:optimaldecoder} satisfies the perception constraint $W^2_2(p_Z,p_{\hat{Z}})\leq P$ and attains the lower bound in \eqref{eq:DPlowerbound}.
	For $P\geq W^2_2(p_Z,p_{\tilde{Z}})$,
	\begin{align}
		W^2_2(p_Z,p_{\hat{Z}})=W^2_2(p_Z,p_{\tilde{Z}})\leq P.\label{eq:perception1}
	\end{align} 
	For $P\in[0,W^2_2(p_Z,p_{\tilde{Z}}))$, 
	\begin{align}
		W^2_2(p_Z,p_{\hat{Z}})&\leq\mathbb{E}[\|Z'-\hat{Z}\|^2]\nonumber\\
		&=\mathbb{E}\left[\left\|\frac{\sqrt{P}}{W_2(p_Z,p_{\tilde{Z}})}(Z'-\tilde{Z})\right\|^2\right]\nonumber\\
		&=\frac{P}{W^2_2(p_Z,p_{\tilde{Z}})}\mathbb{E}[\|Z'-\tilde{Z}\|^2]\nonumber\\
		&=P.\label{eq:perception2}
	\end{align}
	Finally, we have
	\begin{align}
		\mathbb{E}[\|Z-\hat{Z}\|^2]&=\mathbb{E}\left[\left\|(Z-\tilde{Z})-\left(1-\frac{\sqrt{P}}{W_2(p_Z,p_{\tilde{Z}})}\right)_+(Z'-\tilde{Z})\right\|^2\right]\nonumber\\
		&=\mathbb{E}[\|Z-\tilde{Z}\|^2]+\left(1-\frac{\sqrt{P}}{W_2(p_Z,p_{\tilde{Z}})}\right)_+^2\mathbb{E}[\|Z'-\tilde{Z}\|^2]\nonumber\\
		&=\mathbb{E}[\|Z-\tilde{Z}\|^2]+(W_2(p_Z,p_{\tilde{Z}})-\sqrt{P})^2_+.
	\end{align}
%	thereby attaining the lower bound in \eqref{eq:DPlowerbound}. It remains to verify that $p_{\hat{Z}}$ satisfies the perception constraint. Indeed, for $P\geq W^2_2(p_Z,p_{\tilde{Z}})$,
%	\begin{align}
%		W^2_2(p_Z,p_{\hat{Z}})=W^2_2(p_Z,p_{\tilde{Z}})\leq P,
%	\end{align} 
%	while for $P\in[0,W^2_2(p_Z,p_{\tilde{Z}}))$,
%	\begin{align}
%		W^2_2(p_Z,p_{\hat{Z}})&\leq\mathbb{E}[\|Z'-\hat{Z}\|^2]\nonumber\\
%		&=\mathbb{E}\left[\left\|\frac{\sqrt{P}}{W_2(p_Z,p_{\tilde{Z}})}(Z'-\tilde{Z})\right\|^2\right]\nonumber\\
%		&=\frac{P}{W^2_2(p_Z,p_{\tilde{Z}})}\mathbb{E}[\|Z'-\tilde{Z}\|^2]\nonumber\\
%		&=P.
%	\end{align}
\end{remark}

	%In other words, $\hat{S}^*$ is a linear combination of the MMSE estimate $\tilde{S}$ and a perceptually perfect reconstruction $S'$ generated from $\tilde{S}$ 
%through optimal transport. It achieves the best possible distortion-perception tradeoff for the given representation $W$
%since
%\begin{align}
%	W^2_2(p_S,p_{\hat{S}^*})=W^2_2(p_S,p_{\tilde{S}})\wedge P\leq P\label{eq:S*1}
%\end{align}
%and
%\begin{align}
%	\mathbb{E}[\|S-\hat{S}^*\|^2]=\mathbb{E}[\|S-\tilde{S}\|^2]+[(W_2(p_S,p_{\tilde{S}})-\sqrt{P})_+]^2.\label{eq:S*2}
%\end{align}

%add a figure for geometric interpretation

\begin{figure}[htbp]
	\centerline{\includegraphics[width=8.5cm]{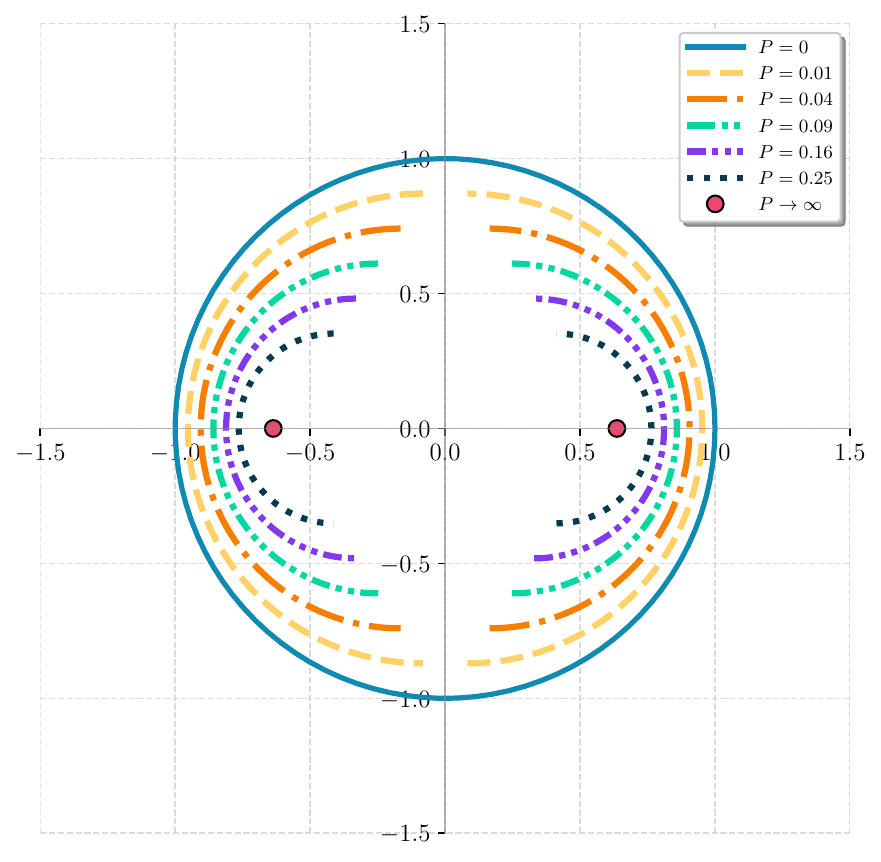}} \caption{Illustration of the supports of $\hat{X}$ across different values of $P$ in the unit-circle example for $1$-bit quantization. Note that $\hat{X}$ coincides with $X'$ when $P=0$, and  with $\tilde{X}$ when $P\rightarrow\infty$ (more precisely, when $P\geq 1-\frac{4}{\pi^2}\approx 0.595$).
    }\label{fig:P}
\end{figure}

It follows from Proposition \ref{prop:interpolation} that choosing 
\begin{align}
	\alpha:=\begin{cases}
		\frac{\sqrt{P}}{W_2(p_X,p_{\tilde{X}})}, &P\in[0,W^2_2(p_X,p_{\tilde{X}})),\\
		1,& P\geq W^2_2(p_X,p_{\tilde{X}}),
	\end{cases}
\end{align}
yields the minimum achievable distortion subject to the perception constraint $P$:
\begin{align}
	\mathbb{E}[\|X-\hat{X}\|^2]&=\mathbb{E}[\|X-\tilde{X}\|^2]+[(W_2(p_X,p_{\tilde{X}})-\sqrt{P})_+]^2\nonumber\\
	&=1-\frac{4}{\pi^2}+\left[\left(\sqrt{1-\frac{4}{\pi^2}}-\sqrt{P}\right)_+\right]^2.\label{eq:tradeoff_ncr}
\end{align}
The random variable  $X'$ is uniformly distributed  over the right unit semicircle when  $\tilde{X}=(\frac{2}{\pi},0)$, and over the left unit semicircle when $\tilde{X}=(-\frac{2}{\pi},0)$. Consequently, for the nondegenerate case $P\in[0,W^2_2(p_X,p_{\tilde{X}}))$, the reconstruction $\hat{X}$ is uniformly distributed over two semicircles of radius $1-\frac{\sqrt{P}}{W_2(p_X,p_{\tilde{X}})}$: specifically, 
the right semicircle  centered at $\frac{\sqrt{P}}{W_2(p_X,p_{\tilde{X}})}(\frac{2}{\pi},0)$ and the left semicircle centered at $\frac{\sqrt{P}}{W_2(p_X,p_{\tilde{X}})}(-\frac{2}{\pi},0)$
 (see Figs. \ref{fig:2*1} and \ref{fig:P}). 
 
It is instructive to interpret the unit circle as an abstract manifold of natural images. In the $1$-bit quantization scenario, minimizing the mean squared error requires the reconstruction 
$\tilde{X}$
to lie strictly inside the circle, rather than on it. In the context of image compression, this corresponds to the fact that optimizing PSNR alone typically drives reconstructed images away from the natural image manifold, resulting in outputs that no longer appear natural. This deviation from the manifold is precisely the source of perception loss. The role of a perception constraint is to enforce proximity between the reconstruction distribution and the source distribution, thereby pushing the reconstruction back toward the unit circle (or the image manifold). In the limiting case of a perfect perception constraint,
$W^2_2(p_X,p_{\hat{X}})=0$, the reconstruction is constrained to lie exactly on the unit circle (or the image manifold), ensuring that it remains a natural image.

%{\bf add a figure for linear interpolation}

%Let $S'$ be a perceptually perfect reconstruction generated from $\tilde{S}$ according to the optimal transport plan that achieves $W_2(p_S,p_{\tilde{S}})$. We construct a reconstruction $\hat{S}$ corresponding to a target perception loss $P$ via the following interpolation scheme:
%\begin{align}
%	\hat{S}:=\begin{cases}
%		\left(1-\frac{\sqrt{P}}{W_2(p_S,p_{\tilde{S}})}\right)S'+\frac{\sqrt{P}}{W_2(p_S,p_{\tilde{S}})}\tilde{S},& P\in[0,W^2_2(p_S,p_{\tilde{S}})),\\
%		\tilde{S},&P\geq W^2_2(p_S,p_{\tilde{S}}).\label{eq:interplation}
%	\end{cases}
%\end{align}
%For the nondegenerate case $P\in[0,W^2_2(p_S,p_{\tilde{S}}))$, given $\tilde{S}$, the distribution of $S'$ is uniform over the arc of the unit circle with the angular range $[-\frac{\pi}{2}+J\pi,\frac{\pi}{2}+J\pi)$ (a semicircle in this case), and consequently $\hat{S}$ is uniformly distributed over an arc of a circle with radius $1-\frac{\sqrt{P}}{W_2(p_S,p_{\tilde{S}})}$, centered at $\frac{\sqrt{P}}{W_2(p_S,p_{\tilde{S}})}\tilde{S}$, with the same angular range. 
%Invoking the fact that
%\begin{align*}
%	\mathbb{E}[\|S-\hat{S}\|^2]=\mathbb{E}[\|S-\tilde{S}\|^2]+[(W_2(p_S,p_{\tilde{S}})-\sqrt{P})_+]^2
%\end{align*}
%yields the following distortion-perception tradeoff:
%\begin{align*}
%	D=1-\frac{4}{\pi^2}+\left[\left(\sqrt{1-\frac{4}{\pi^2}}-\sqrt{P}\right)_+\right]^2,
%\end{align*}
%where $D:=\mathbb{E}[\|S-\hat{S}\|^2]$.

\subsection{$1$-bit Quantization with $1$-bit of Common Randomness}

We now introduce an additional resource---common randomness---in the form of a random seed 
$K$ shared between the encoder and decoder. At first glance, it may seem surprising that common randomness can be useful at all. Indeed, in conventional lossy source coding, restricting both the encoder and decoder to be deterministic incurs no loss of optimality. In contrast, under a perception constraint, the decoder typically must be stochastic in order to fulfill the associated generative task. Nevertheless, the benefit of common randomness is not immediately apparent, since stochasticity at the decoder could, in principle, be generated locally. To better understand the role of common randomness, we revisit the unit-circle example and consider the setting of $1$-bit quantization with an additional $1$-bit shared random seed.

 %no need to introduce $\Theta$ here

\begin{figure}[htbp]
 	\centerline{\includegraphics[width=18cm]{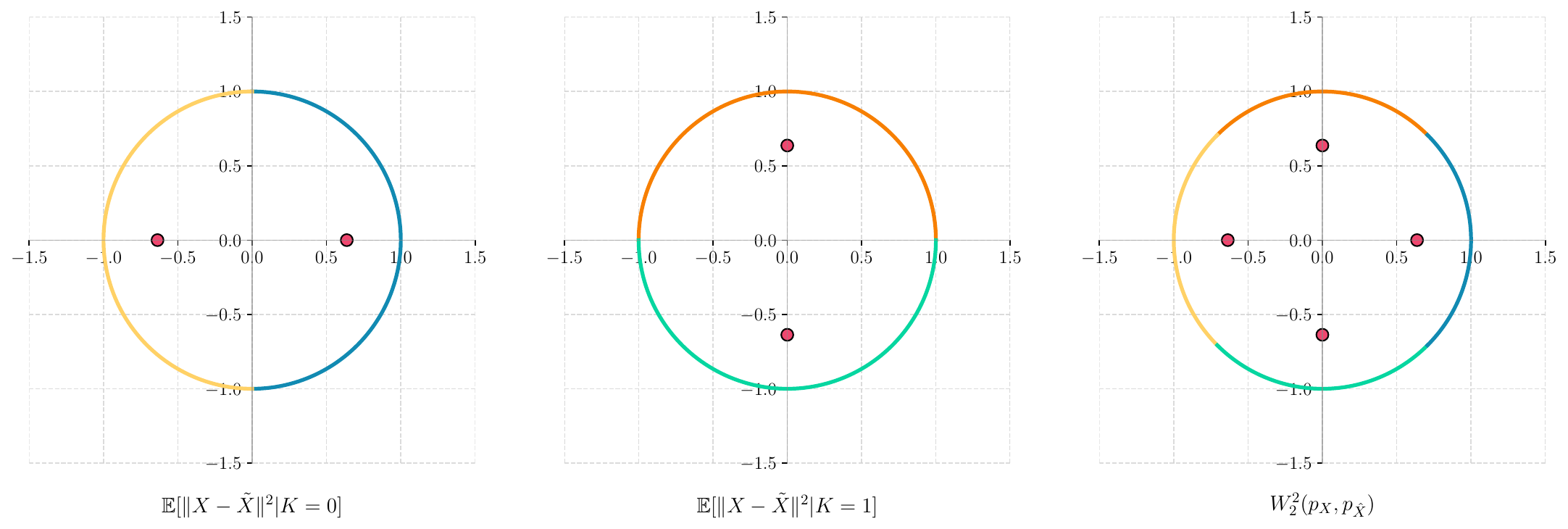}} \caption{
 Illustration of the distinction between the computations of
 $\mathbb{E}[\|X-\tilde{X}\|^2]$ and $W^2_2(p_X,p_{\hat{X}})$ in the unit-circle example
 for $1$-bit quantization with $1$-bit of common randomness. }\label{fig:operational_vs_virtual}
 \end{figure}

Let $K$ be a random seed uniformly distributed over $\{0,1\}$ and independent of $X$. Define the encoder $f:\mathbb{S}\times\{0,1\}\rightarrow\{0,1\}$ as follows: when $K=0$, set $J:=f(X,K)=0$ if $X$ lies on the right unit semicircle and $J=1$ if $X$ lies on the left unit semicircle; when $K=1$, set $J=0$ when $X$ lies on the upper unit semicircle and $J=1$ if $X$ lies on the lower unit semicircle.
Moreover, let $\tilde{X}:=\mathbb{E}[X|J,K]$. It can be verified that $\tilde{X}$ is uniformly distributed over the centroids of the four unit semicircles, namely,
\begin{align}
	\tilde{X}=\begin{cases}
		(\frac{2}{\pi},0),& J=0, K=0,\\
		(0,\frac{2}{\pi}),&J=0,K=1,\\
		(-\frac{2}{\pi},0),&J=1,K=0,\\
		(0,-\frac{2}{\pi}),& J=1,K=1.
	\end{cases}
\end{align}
%When $K=0$, the scheme reduces to the aforementioned $1$-bit quantization  mapping the right and left unit semicircles to their  centroids $(\frac{2}{\pi},0)$ and $(-\frac{2}{\pi},0)$, respectively, while for $K=1$, it maps the upper and lower semicircles to $(0,\frac{2}{\pi})$ and $(0,-\frac{2}{\pi})$. 
By symmetry (see Fig. \ref{fig:operational_vs_virtual}),
\begin{align}
	\mathbb{E}[\|X-\tilde{X}\|^2]&=\frac{1}{2}\mathbb{E}[\|X-\tilde{X}\|^2|K=0]+\frac{1}{2}\mathbb{E}[\|X-\tilde{X}\|^2|K=1]\nonumber\\
	&=\mathbb{E}[\|X-\tilde{X}\|^2|K=0]\nonumber\\
	&=1-\frac{4}{\pi^2},
\end{align}
identical to the distortion loss in \eqref{eq:distortion_1bit}. However, the induced $\tilde{X}$ is now supported on four centroids, and the associated minimum-distance transport plan partitions the unit circle into four equal arcs (see Fig. \ref{fig:operational_vs_virtual}).
Consequently, the perception loss is given by
\begin{align}
	W^2_2(p_X,p_{\tilde{X}})&=\frac{2}{\pi}\int_{-\frac{\pi}{4}}^{\frac{\pi}{4}}(\frac{2}{\pi}-\cos(\theta))^2+\sin^2(\theta)\mathrm{d}\theta\nonumber\\
	&=1-\frac{4(2\sqrt{2}-1)}{\pi^2},
\end{align}
which is strictly smaller than that in \eqref{eq:perception_1bit}. Through linear interpolation of the MMSE estimate $\tilde{X}$ with a perceptually perfect sample $X'$ (generated from $\tilde{X}$ via a transport plan attaining $W^2_2(p_X,p_{\tilde{X}})$), one obtains a reconstruction 
 $\hat{X}$
that yields 
\begin{align}
	\mathbb{E}[\|X-\hat{X}\|^2]=1-\frac{4}{\pi^2}+\left[\left(\sqrt{1-\frac{4(2\sqrt{2}-1)}{\pi^2}}-\sqrt{P}\right)_+\right]^2,\label{eq:tradeoff_1bit}
\end{align}
thereby improving  the  distortion-perception tradeoff in \eqref{eq:tradeoff_ncr} for $1$-bit quantization without common randomness.

%Next consider the case with 1-bit quantization and 1-bit common randomness, i.e., $M=N=2$. Let 

\subsection{Extension}

The above scheme readily extends to the setting where neither the encoder output nor the shared random seed is restricted to be binary.
Specifically, let  $K$ be uniformly distributed over $\mathcal{K}$ and independent of $X$, and define $J:=f(X,K)$, where $f:\mathbb{S}\times\mathcal{K}\rightarrow\mathcal{J}$ is a (possibly stochastic) encoder. We assume $\mathcal{J}:=\{0,1,\ldots,M-1\}$ and $\mathcal{K}:=\{0,1,\ldots,N-1\}$ , with $M$ and $N$ being positive integers.

%, $J$ be the encoder output, and $K$ be a shared random seed. Let $\Theta$ denote the angle of $S$, with all angles interpreted modulo $2\pi$, i.e., identified up to integer multiples of $2\pi$. We assume $\mathcal{J}:=\{0,1,\ldots,M-1\}$ and $\mathcal{K}:=\{0,1,\ldots,N-1\}$. We adopt the squared-error distortion measure and the squared Wasserstein-2 perception measure. 

%Writing $\Theta$ for the angle of $X$ (interpreted modulo 
%$2\pi$), this corresponds to assigning
%$J=0$ for $\Theta\in[-\frac{\pi}{2},\frac{\pi}{2})$ and $J=1$ for $\Theta\in[\frac{\pi}{2},\frac{3\pi}{2})$%.

%Finally consider the general case with $M\geq 1$ and $N\geq 1$. Let
%$J:=j$ if $\Theta\in[\frac{((2j-1)N+2K)\pi}{MN},\frac{((2j+1)N+2K)\pi}{MN})$ for $j\in\mathcal{J}$. 

For each realization of $K$, partition the unit circle into $M$ equal arcs and assign distinct encoder outputs to different arcs, with different values of $K$ corresponding to rotations of this partition.
Specifically, select $J\in\mathcal{J}$ such that
\begin{align}
\Theta\in\left[\frac{((2j-1)N+2K)\pi}{MN},\frac{((2j+1)N+2K)\pi}{MN}\right),\label{eq:encoding}
\end{align}
where $\Theta$ denotes the angle of $X$, with all angles interpreted modulo $2\pi$, i.e., identified up to integer multiples of $2\pi$.
It can be verified that
\begin{align}
	\tilde{X}&:=\mathbb{E}[X|J,K]\nonumber\\
	&=\frac{M\sin(\frac{\pi}{M})}{\pi}\left(\cos\left(\frac{2(JN+K)\pi}{MN}\right),\sin\left(\frac{2(JN+K)\pi}{MN}\right)\right).\label{eq:decoding}
\end{align}
The corresponding distortion and perception losses are given by
\begin{align}
	\mathbb{E}[\|X-\tilde{X}\|^2]&=\frac{M}{2\pi}\int_{-\frac{\pi}{M}}^{\frac{\pi}{M}}\left(\frac{M\sin(\frac{\pi}{M})}{\pi}-\cos(\theta)\right)^2+\sin^2(\theta)\mathrm{d}\theta\nonumber\\
	&=1-\frac{M^2\sin^2(\frac{\pi}{M})}{\pi^2}
\end{align}
and
\begin{align}
	W^2_2(p_X,p_{\tilde{X}})&=\frac{MN}{2\pi}\int_{-\frac{\pi}{MN}}^{\frac{\pi}{MN}}\left(\frac{M\sin(\frac{\pi}{M})}{\pi}-\cos(\theta)\right)^2+\sin^2(\theta)\mathrm{d}\theta\nonumber\\
	&=1-\frac{M^2\left(2N\sin(\frac{\pi}{MN})-\sin(\frac{\pi}{M})\right)\sin(\frac{\pi}{M})}{\pi^2},
\end{align}
respectively.
Construct $\hat{X}$ using the linear interpolation scheme yields the following distortion-perception tradeoff:
\begin{align}
	\mathbb{E}[\|X-\hat{X}\|^2]=1-\frac{M^2\sin^2(\frac{\pi}{M})}{\pi^2}+\left[\left(\sqrt{1-\frac{M^2\left(2N\sin(\frac{\pi}{MN})-\sin(\frac{\pi}{M})\right)\sin(\frac{\pi}{M})}{\pi^2}}-\sqrt{P}\right)_+\right]^2.\label{eq:general}
\end{align}
For the nondegenerate case $P\in[0,W^2_2(p_X,p_{\tilde{X}}))$, conditioned on $\tilde{X}$, the reconstruction $\hat{X}$ is uniformly distributed over an arc of radius $1-\frac{\sqrt{P}}{W_2(p_X,p_{\tilde{X}})}$, centered at $\frac{\sqrt{P}}{W_2(p_X,p_{\tilde{X}})}\tilde{X}$, with  angular range $[\frac{(2JN+2K-1)\pi}{MN},\frac{(2JN+2K+1)\pi}{MN})$.

%Setting $M=2$ and $N=\infty$ in \eqref{eq:general}

%reduces to dithering

%the distribution of $S'$ is uniform over the arc of the unit circle with the angular range $[-\frac{\pi}{2}+J\pi,\frac{\pi}{2}+J\pi)$ (a semicircle in this case), and consequently $\hat{S}$ is uniformly distributed over an arc of a circle with radius $\frac{\sqrt{P}}{W_2(p_S,p_{\tilde{S}})}$, centered at $(1-\frac{\sqrt{P}}{W_2(p_S,p_{\tilde{S}})})\tilde{S}$, with the same angular range.

Note that \eqref{eq:general} reduces to
\begin{align}
\mathbb{E}[\|X-\hat{X}\|^2]=1-\frac{M^2\sin^2(\frac{\pi}{M})}{\pi^2}+\left[\left(\sqrt{1-\frac{M^2\sin^2(\frac{\pi}{M})}{\pi^2}}-\sqrt{P}\right)_+\right]^2\label{eq:N=1}
\end{align}
when $N=1$ (no common randomness), and to 
\begin{align}
\mathbb{E}[\|X-\hat{X}\|^2]=1-\frac{M^2\sin^2(\frac{\pi}{M})}{\pi^2}+\left[\left(\sqrt{1-\frac{2M\sin(\frac{\pi}{M})}{\pi}+\frac{M^2\sin^2(\frac{\pi}{M})}{\pi^2}}-\sqrt{P}\right)_+\right]^2\label{eq:N=infty}
\end{align}
as $N\rightarrow\infty$ (unlimited common randomness). Setting $P=0$ in \eqref{eq:N=1} and \eqref{eq:N=infty} further yields
\begin{align}
	\mathbb{E}[\|X-\hat{X}\|^2]=2-\frac{2M^2\sin^2(\frac{\pi}{M})}{\pi^2}
\end{align}
and
\begin{align}
	\mathbb{E}[\|X-\hat{X}\|^2]=2-\frac{2M\sin(\pi/M)}{\pi},
\end{align}
respectively, thereby recovering \cite[Eqs. (50) and (51)]{TA21}. %In the limit case $N\rightarrow\infty$ and $P=0$, the stochastic mapping from $X$ to $\hat{X}$ admits a simple interpretation \cite{TA21} as dithered quantization \cite{Ziv85}. \textcolor{blue}{Three questions here: 1) Should we reference Zamir's paper as well? 2) I feel it'll help to mention that when $N=\infty$, a simple strategy to achieve the tradeoff is to interpolate between dithered quantization and the MSE estimator? We want to make it more explicit such constructions 3) The dithering may need some clarification here. It needs to be uniformly distributed on certain cell, and here it is on the circle, instead of the general 2D space.}

%In this limiting case, the mapping from $X$ to $\tilde{X}$ can be interpreted as dithered quantization.

\begin{figure}[htbp]
	\centerline{\includegraphics[width=8.5cm]{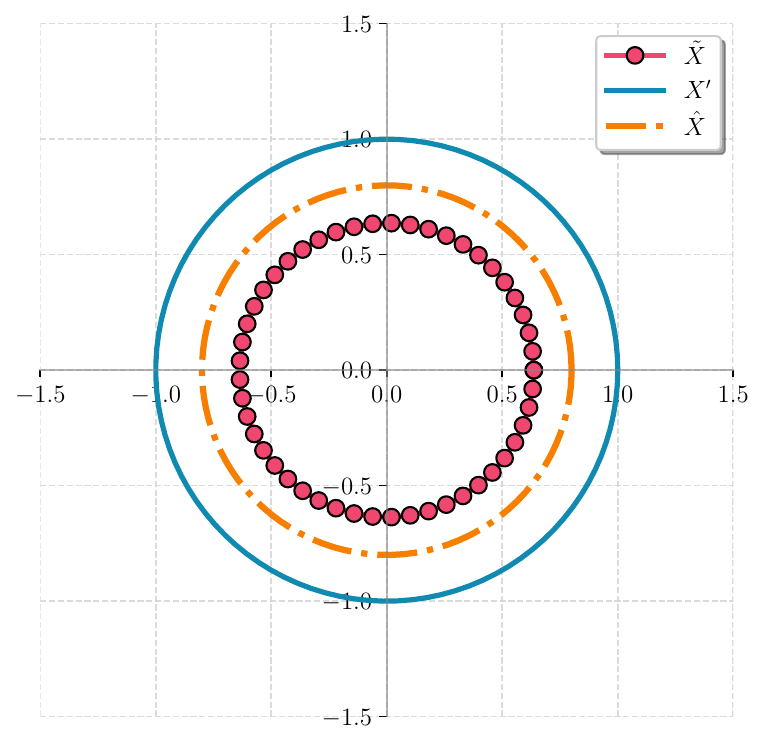}} \caption{Illustration of the supports of $\tilde{X}$, $X'$, and $\hat{X}$ (with $P=0.04$) in the unit-circle example for  $1$-bit quantization with unlimited common randomness. In this case, $\tilde{X}$, $X'$, and $\hat{X}$ are  uniformly distributed over circles centered at the origin with radii $\frac{2}{\pi}\approx0.637$, $1$, and $0.8$, respectively, and are deterministically related through scaling.
    }\label{fig:N=infty}
\end{figure}

%need to correct $P$

In the large-$N$ limit, $\tilde{X}$, $X'$, and $\hat{X}$ are uniformly distributed over circles centered at the origin with radii with $\frac{M\sin(\frac{\pi}{M})}{\pi}$, $1$, and
\begin{align}
\min\left\{\sqrt{\frac{P\pi^2}{\pi^2-M^2(\frac{2\pi}{M}-\sin(\frac{\pi}{M}))\sin(\frac{\pi}{M})}},1\right\}\left(\frac{M\sin(\frac{\pi}{M})}{\pi}-1\right)+1,
\end{align}
respectively, and are   deterministically related through scaling (see Fig. \ref{fig:N=infty}). Moreover, the construction of $\tilde{X}$ (equivalently, $X'$ and $\hat{X}$) can be realized via dithered quantization \cite{Ziv85}.
Indeed, \eqref{eq:encoding} effectively implements the mapping
\begin{align}
\Theta-\frac{2K\pi}{MN}\rightarrow\frac{2J\pi}{M},
\end{align}
where the index $J$ satisfies
\begin{align}
\Theta-\frac{2K\pi}{MN}\in\left[\frac{(2J-1)\pi}{M},\frac{(2J+1)\pi}{M}\right),\label{eq:dither1}
\end{align}
while \eqref{eq:decoding} can be decomposed into the mappings
\begin{align}
\frac{2J\pi}{M}\rightarrow\tilde{\Theta}:=\frac{2J\pi}{M}+\frac{2K\pi}{MN}\label{eq:dither2}
\end{align}
and
\begin{align}
\tilde{\Theta}\rightarrow\tilde{X}:=\frac{M\sin(\frac{\pi}{M})}{\pi}(\cos(\tilde{\Theta}),\sin(\tilde{\Theta})).
\end{align}
As $N\rightarrow\infty$, the term $\frac{2K\pi}{MN}$ becomes uniformly distributed over $[0,\frac{2\pi}{M})$; consequently, the mappings \eqref{eq:dither1} and \eqref{eq:dither2} reduce to conventional dithered quantization performed in the angular domain.

%\begin{align}
%\tilde{X}:=\frac{M\sin(\frac{\pi}{M})}{\pi}(\cos(\tilde{\Theta}),\sin(\tilde{\Theta})),
%\end{align}
%where
%\begin{align}
%\tilde{\Theta}:=\frac{2J\pi}{M}+\frac{2K\pi}{MN}.
%\end{align}

%$\hat{X}$ is uniformly distributed over a circle of radius 
%\begin{align}
%\sqrt{\frac{P\pi^2}{\pi^2-M^2(\frac{2\pi}{M}-\sin(\frac{\pi}{M}))\sin(\frac{\pi}{M})}}\left(\frac{M\sin(\frac{\pi}{M})}{\pi}-1\right)+1,
%\end{align}
%centered at the origin; moreover, $\tilde{X}$, $X'$, and $\hat{X}$ become scaled versions of each other (see Fig. \ref{fig:N=infty}).
%As a consequence, one can 
%by bypass the steps of estimating $\tilde{X}$ and generating $X'$ and directly construct $\hat{X}$. It turns out that this can be accomplished via dithered quantization. 

\begin{figure}[htbp]
	\centerline{\includegraphics[width=12cm]{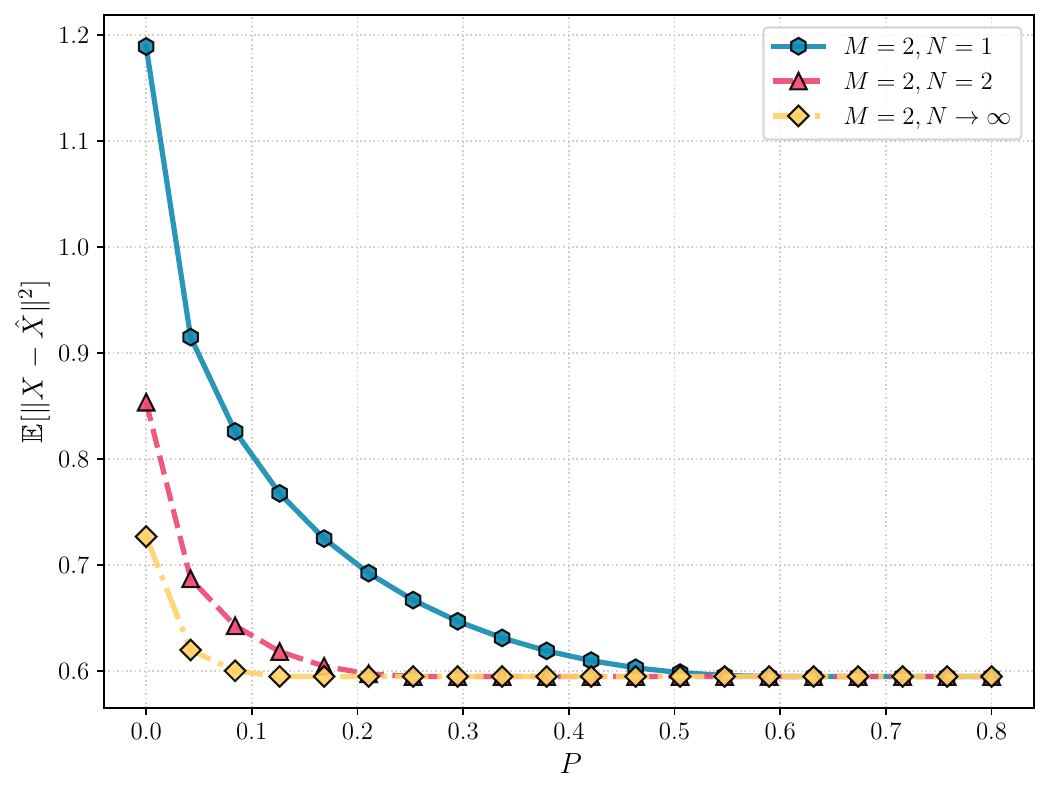}} \caption{Comparison of the distortion–perception tradeoffs in the unit-circle example for  $1$-bit quantization  with  no common randomness,  $1$-bit of common randomness, and  unlimited common randomness.}\label{fig:differentN}
\end{figure}

To gain a concrete understanding of the impact of common randomness on the distortion-perception tradeoff, we plot 
\eqref{eq:general} for the case $M=2$ ($1$-bit quantization) with $N=1$ (no common randomness), $N=2$ ($1$-bit of common randomness), and $N\rightarrow\infty$ (unlimited common randomness),  corresponding, respectively, to \eqref{eq:tradeoff_ncr}, \eqref{eq:tradeoff_1bit}, and
\begin{align}
\mathbb{E}[\|X-\hat{X}\|^2]=1-\frac{4}{\pi^2}+\left[\left(\sqrt{1-\frac{4}{\pi}+\frac{4}{\pi^2}}-\sqrt{P}\right)_+\right]^2.
\end{align}
It can be seen from Fig.~\ref{fig:differentN} that increasing the amount of common randomness improves the distortion–perception tradeoff, particularly in the near-perfect perception regime.

%{\bf plot the D-P tradeoff for $M=2, N=1$ and $M=2, N=2$, $M=2, N=\infty$}

\begin{figure}[htbp]
	\centerline{\includegraphics[width=18cm]{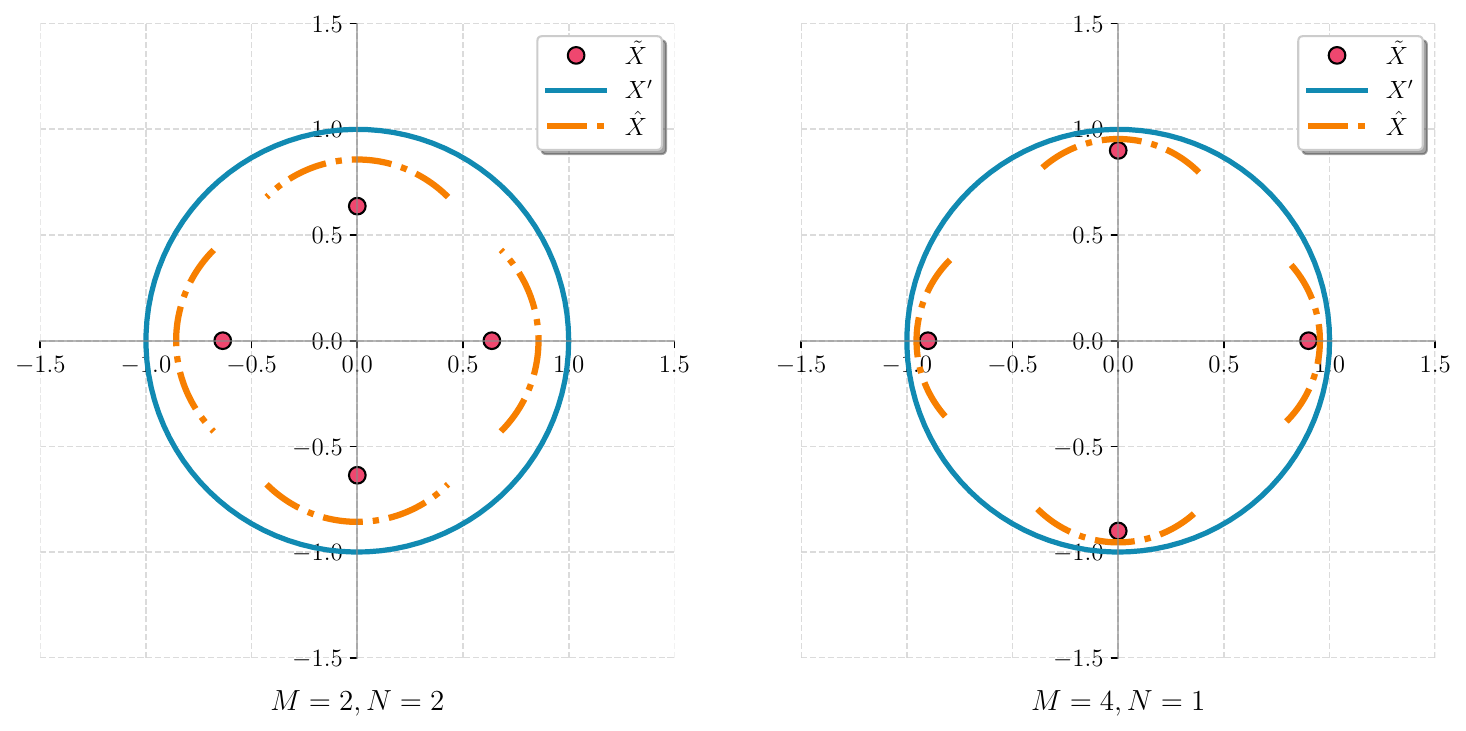}} \caption{Comparison of the supports of $\tilde{X}$, $X'$, and $\tilde{X}$ (with $P=0.04$) in the unit-circle example for $1$-bit quantization  with $1$-bit of common randomness ($M=2$, $N=2$) and $2$-bit quantization with no common randomness ($M=4$, $N=1$). Specifically, in the first setting, $\tilde{X}$ is uniformly distributed over the four points $(\pm\frac{2}{\pi},0)\approx(\pm 0.637, 0)$ and $(0,\pm\frac{2}{\pi})\approx(0,\pm 0.637)$, while $\hat{X}$ is uniformly distributed over four quarter circles of radius $1-\frac{\pi}{5\sqrt{\pi^2-4(2\sqrt{2}-1)}}\approx0.607$, centered at $(\pm\frac{2}{5\sqrt{\pi^2-4(2\sqrt{2}-1)}},0)\approx(\pm 0.250,0)$ and $(0,\pm\frac{2}{5\sqrt{\pi^2-4(2\sqrt{2}-1)}})\approx(0,\pm 0.250)$, respectively; in the second setting, $\tilde{X}$ is uniformly distributed over the four points $(\pm\frac{2\sqrt{2}}{\pi},0)\approx(\pm 0.900, 0)$ and $(0,\pm\frac{2\sqrt{2}}{\pi})\approx(0,\pm 0.900)$, while $\hat{X}$ is uniformly distributed over four quarter circles of radius $1-\frac{\pi}{5\sqrt{\pi^2-8}}\approx0.540$, centered at $(\pm\frac{2\sqrt{2}}{5\sqrt{\pi^2-8}},0)\approx(\pm 0.414,0)$ and $(0,\pm\frac{2\sqrt{2}}{5\sqrt{\pi^2-8}})\approx(0,\pm 0.414)$, respectively; in both settings, $X'$ is uniformly distributed over the unit circle centered at the origin.
}\label{fig:2*2vs4*1}
\end{figure}

\begin{figure}[htbp]
	\centerline{\includegraphics[width=12cm]{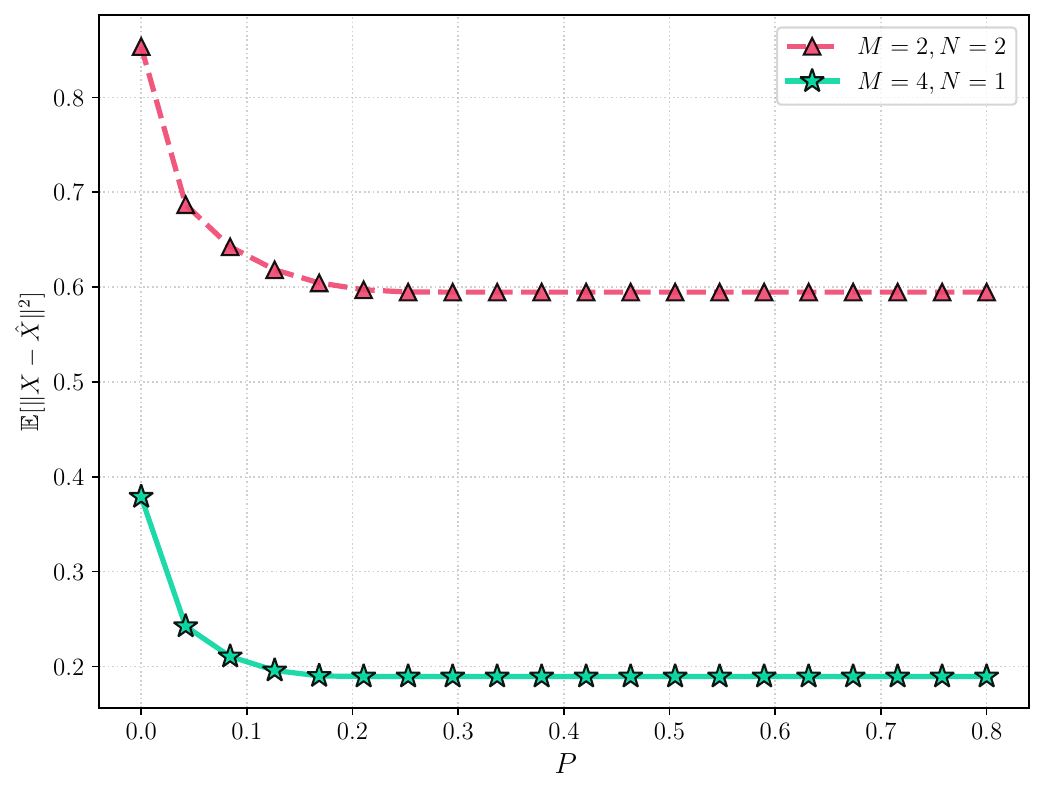}} \caption{Comparison of the distortion-perception tradeoffs in the unit-circle example for $1$-bit quantization  with $1$-bit of common randomness ($M=2$, $N=2$) and $2$-bit quantization with no common randomness ($M=4$, $N=1$) .}\label{fig:info_vs_common}
\end{figure}

We close this section by highlighting the distinction between common randomness bits and information bits. To this end, consider the cases of $M=2$, $N=2$ ($1$-bit quantization with $1$-bit of common randomness) and $M=4$, $N=1$ ($2$-bit quantization without common randomness). As illustrated in Fig. \ref{fig:2*2vs4*1}, the corresponding reconstruction distributions are qualitatively different. This difference is also reflected in their associated distortion–perception tradeoffs,  given by \eqref{eq:tradeoff_1bit} and
\begin{align}
	\mathbb{E}[\|X-\hat{X}\|^2]=1-\frac{8}{\pi^2}+\left[\left(\sqrt{1-\frac{8}{\pi^2}}-\sqrt{P}\right)_+\right]^2,\label{eq:M=4,N=1}
\end{align}
respectively. As seen from Fig. \ref{fig:info_vs_common},  the latter strictly outperforms the former. This disparity can be understood through the general tradeoff expression
\begin{align}
	\mathbb{E}[\|X-\hat{X}\|^2]=\mathbb{E}[\|X-\tilde{X}\|^2]+[(W_2(p_X,p_{\tilde{X}})-\sqrt{P})_+]^2.
\end{align}
Increasing the amount of common randomness   reduces only $W^2_2(p_X,p_{\tilde{X}})$, whereas increasing the number of information bits reduces both  $\mathbb{E}[\|X-\tilde{X}\|^2]$ and  $W^2_2(p_X,p_{\tilde{X}})$. %\textcolor{green}{It might be nice if we can include a different value of $P$ to illustrate the "trend", and explain that when $P=0$ they collapse to the same, but via different mechanism.}

\section{The One-Shot Setting}\label{sec:one-shot}

\subsection{System Architecture}

\begin{figure}[htbp]
	\centerline{\includegraphics[width=21cm]{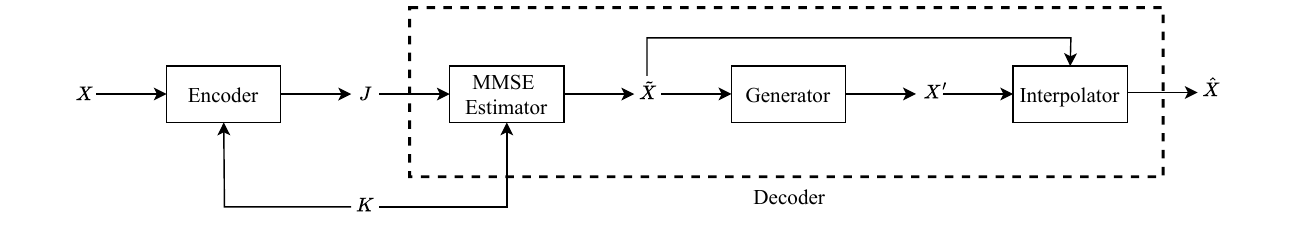}} \caption{
		System diagram of perception-aware lossy source coding. }\label{fig:system_diagram}
\end{figure}

The architectural insights gained from the unit-circle example extend naturally to general sources. Let $X$ be a random vector in $\mathbb{R}^d$ with $\mathbb{E}[\|X\|^2]<\infty$, and let $K$ be a random seed uniformly distributed over $\mathcal{K}$ and independent of $X$. A perception-aware lossy source coding system (see Fig. \ref{fig:system_diagram}) consists of a (possibly stochastic) encoder $f:\mathrm{R}^d\times\mathcal{K}\rightarrow\mathcal{J}$ that maps $(X,K)$ to an index $J$ and a (possibly stochastic) decoder $g:\mathcal{J}\times\mathcal{K}\rightarrow\mathbb{R}^d$ that generates a reconstruction $\hat{X}$ based on $(J,K)$. Given the encoder, the optimal decoder
comprises an MMSE estimator that computes
 $\mathbb{E}[X|J,K]$, a generator that produces a perceptually perfect sample
 $X'$ from $\tilde{X}$ via a transport plan attaining  $W^2_2(p_X,p_{\tilde{X}})$, and an interpolator that forms the reconstruction $\hat{X}:=(1-\alpha)X'+\alpha\tilde{X}$ as a convex combination of $X'$ and $\tilde{X}$,
 where
\begin{align}
	\alpha:=\begin{cases}
		\frac{\sqrt{P}}{W_2(p_X,p_{\tilde{X}})}, &P\in[0,W^2_2(p_X,p_{\tilde{X}})),\\
		1,& P\geq W^2_2(p_X,p_{\tilde{X}}).
	\end{cases}\label{eq:alpha}
\end{align}
We assume $\mathcal{J}:=\{0,1,\ldots,M-1\}$ and $\mathcal{K}:=\{0,1,\ldots,N-1\}$, where $M$ and $N$ are fixed positive integers. With  the optimal decoder  fully specified,  the minimum achievable distortion with respect to a fixed encoder, subject to the perception constraint
 $W^2_2(p_X,p_{\hat{X}})\leq P$,  is given by
\begin{align}
	\mathbb{E}[\|X-\hat{X}\|^2]=\mathbb{E}[\|X-\tilde{X}\|^2]+[(W_2(p_X,p_{\tilde{X}})-\sqrt{P})_+]^2.\label{eq:optimaltradeoff}
\end{align}
Consequently, the system design reduces to optimizing the encoder $f$.

 %reconstruction $\hat{X}$ meeting the perception constraint $W^2_2(p_X,p_{\hat{X}})\leq P$ using a convex combination of $\tilde{X}$ and $X'$. 

%$\mathcal{J}:=\{0,1,\ldots,M-1\}$ and $\mathcal{K}:=\{0,1,\ldots,N-1\}$

In practice, the encoder is typically implemented using a neural network followed by a quantization module. The neural network maps the input $X$, conditioned on the realization of the shared random seed $K$, into a lower-dimensional latent representation. This reduction  is often well justified, as in many practical scenarios the pristine signal resides on a lower-dimensional manifold embedded in a higher-dimensional ambient space. A simple illustration is provided by the unit-circle example, where the source variable is confined to a $1$-dimensional circle in a $2$-dimensional Euclidean space. In such cases, the latent representation can be interpreted as a reparameterization of the underlying signal manifold that captures its intrinsic dimensionality; for instance, expressing points on the unit circle in polar coordinates reduces the representation to a single parameter, namely the angle. By aligning the representation with the intrinsic structure of the data, this dimensionality reduction enables more efficient quantization, as it avoids redundancy in the ambient space and focuses resolution on the most relevant degrees of freedom.

The quantization module then converts this continuous representation into a discrete form. The most common choice is scalar uniform quantization applied element-wise, due to its simplicity, although vector quantization or learned codebooks can also be employed to achieve higher compression efficiency at the cost of increased complexity. A key challenge here is that quantization is inherently non-differentiable, which complicates end-to-end training via gradient-based methods. To address this issue, neural compression approaches typically employ probabilistic or continuous relaxations---such as additive noise or soft quantization---to approximate hard quantization during training while retaining differentiability.

The MMSE estimator can also be implemented using a neural network, which can be optimized via supervised training for a fixed encoder. In contrast, the design of the generator that produces $X'$ from $\tilde{X}$ is more involved, as it requires identifying a transport plan that attains
 $W^2_2(p_X,p_{\tilde{X}})$ and training a generative model to realize this plan.
 A simple but potentially suboptimal alternative is to replace the optimal transport plan with posterior sampling,
 which increases $\mathbb{E}[\|X'-\tilde{X}\|^2]$ from $W^2_2(p_X,p_{\tilde{X}})$ to $\mathbb{E}[\|X-\tilde{X}\|^2]$. In this case, adjusting the interpolation parameter in \eqref{eq:alpha} to
\begin{align}
	\alpha:=\begin{cases}
		\sqrt{\frac{P}{\mathbb{E}[\|X-\tilde{X}\|^2]}}, &P\in\left[0,\sqrt{\mathbb{E}[\|X-\tilde{X}\|^2]}\right),\\
		1,& P\geq \sqrt{\mathbb{E}[\|X-\tilde{X}\|^2]},
	\end{cases}\label{eq:alpha'}
\end{align}
achieves 
\begin{align}
	\mathbb{E}[\|X-\hat{X}\|^2]=\mathbb{E}[\|X-\tilde{X}\|^2]+\left[\left(\sqrt{\mathbb{E}[\|X-\tilde{X}\|^2]}-\sqrt{P}\right)_+\right]^2,\label{eq:suboptimal}
\end{align}
which is generally worse than \eqref{eq:optimaltradeoff}. This gap can be reduced by employing the rectified flow approach \cite{OME25}. Specifically, a rectified flow model \cite{LGL23} is trained using paired samples $(X,\tilde{X})$ to learn a transport map between the two distributions. Leveraging its straight-path property---where trajectories between 
$\tilde{X}$
and 
$X$ are encouraged to follow approximately linear paths in the data space---the resulting model induces a transport plan with transport cost $\mathbb{E}[\|X'-\tilde{X}\|^2]$  no greater than the initial coupling cost $\mathbb{E}[\|X-\tilde{X}\|^2]$. Furthermore, through iterative `Reflow' procedures, this transport cost can be progressively minimized to approach
$W^2_2(p_X,p_{\tilde{X}})$. By adjusting the interpolation parameter to match this improved transport cost, one obtains 
\begin{align}
	\mathbb{E}[\|X-\hat{X}\|]=\mathbb{E}[\|X-\tilde{X}\|^2]+\left[\left(\sqrt{\mathbb{E}[\|X'-\tilde{X}\|^2]}-\sqrt{P}\right)_+\right]^2,\label{eq:improve}
\end{align}
which narrows the gap to optimality.

\subsection{One-Shot Universal Representation}

It is worth noting that optimizing the decoder---specifically, training the MMSE estimator, designing the generator (whether via posterior sampling or rectified flow), and selecting the interpolation parameter---requires a fixed encoder. At the same time, the encoder itself must be optimized, leading to a coupled design problem that complicates the overall training procedure. This difficulty is greatly alleviated in the following scenario.
We say a one-shot universal representation exists if the optimal encoder that minimizes
\begin{align}
	\mathbb{E}[\|X-\tilde{X}\|^2]+[(W_2(p_X,p_{\tilde{X}})-\sqrt{P})_+]^2\label{eq:objective}
\end{align}
does not depend on the perception level $P$. In this case, one can instead consider the limiting objective $\mathbb{E}[\|X-\tilde{X}\|^2]$ obtained by letting $P\rightarrow\infty$ in \eqref{eq:objective}. Consequently, the encoder design decouples from the perception constraint and reduces to that of a conventional, perception-oblivious MMSE quantizer. In other words, the encoder, together with the MMSE estimator, can be optimized solely by minimizing distortion, after which the generator is constructed accordingly; only the interpolator  remains to depend on the target perception level $P$. It follows from 
 Proposition \ref{prop:1} that a one-shot universal representation exists 
in the absence of common randomness (i.e., $N=1$), as the optimal encoder simultaneously minimizes $\mathbb{E}[\|X-\tilde{X}\|^2]$ and $W^2_2(p_X,p_{\tilde{X}})$. Moreover, since
 $W^2_2(p_X,p_{\tilde{X}})$ coincides with $\mathbb{E}[\|X-\hat{X}\|^2]$ in this setting, the optimal transport plan reduces to posterior sampling.  
 
 On the other hand, in the presence of common randomness, there appears to be a potential tension between $\mathbb{E}[\|X-\tilde{X}\|^2]$ and $W^2_2(p_X,p_{\tilde{X}})$, suggesting that a one-shot universal representation may not always exist. The unit-circle example provides useful intuition: owing to its rotational symmetry, one can preserve $\mathbb{E}[\|X-\tilde{X}\|^2]$ while reducing $W^2_2(p_X,p_{\tilde{X}})$ by
 switching, according to the realization of the random seed $K$,  among different MMSE quantizers. For  a generic source, however, such flexibility may be limited. There may not be many MMSE quantizers available, and even when they exist, such quantizers can be too closely aligned to provide sufficient diversity for effectively reducing $W^2_2(p_X,p_{\tilde{X}})$. Consequently, one may need to resort to suboptimal quantizers to decrease $W^2_2(p_X,p_{\tilde{X}})$, at the price of increasing $\mathbb{E}[\|X-\tilde{X}\|^2]$.
A deeper understanding of this tension and its underlying mechanisms is therefore of both theoretical and practical importance. More broadly, similar tradeoffs can be studied under general distortion and perception measures \cite{FWM24}, which may ultimately lead to a unified and comprehensive rate–distortion–perception theory.

 %In general, there appears to exist tension between $\mathbb{E}[\|X-\tilde{X}\|^2]$ and $W^2_2(p_X,p_{\tilde{X}})$, and consequently a one-shot universal representation might not exist.
 %However, both theoretical and empirical evidence suggests that, even when common randomness is available, adopting the MMSE quantizer as the encoder incurs little, if any, performance loss in terms of the distortion–perception tradeoff \cite{ZQCK21}. Nevertheless, a deeper understanding of this tension is of both theoretical and practical significance. In fact, one can even study similar tensions under arbitrary distortion and perception measures \cite{FWM24}, which will pave the way for developing a general rate-distortion-perception theory.
 
 %\textcolor{blue}{For example, for the scalar Gaussian source, such universality indeed holds; in contrast, for vector Gaussian sources, such universality usually does not hold.}
 
 \subsection{Variable-Length Coding}

In this tutorial, we focus on fixed-length coding for simplicity, where the rate is fully specified by the cardinality of $\mathcal{J}$, namely, $R:=\lceil\log N\rceil$. When variable-length coding is employed, additional considerations arise. In particular, one must incorporate learned probabilistic models---often referred to as entropy models---for the latent variables \cite{BLS17}. 
These models enable efficient entropy coding (e.g., via arithmetic coding) by estimating the distribution of the quantized latent representation, frequently utilizing hierarchical priors to capture spatial dependencies \cite{MBI2018}. Accordingly, the training objective is modified to balance rate and distortion, typically taking the form
\begin{align}
	\mathbb{E}[\|X-\hat{X}\|^2]+\lambda R\label{eq:variable_length}
\end{align}
where 
$R$
denotes the expected number of bits required to encode the quantized latent representation, and 
$\lambda$ controls the tradeoff between compression rate and reconstruction fidelity. %\textcolor{green}{Should we mention functional representation here, since it is actually a variable length code.}

With the availability of unlimited common randomness, the reconstruction 
$\hat{X}$ can be expressed as a deterministic function of the source $X$ and the shared random seed 
$K$. This observation allows one to bypass the intermediate steps of forming the MMSE estimate 
$\tilde{X}$
 and generating the perceptually perfect sample 
$X'$, and instead construct $\hat{X}$ directly\footnote{In the unit-circle example, this simplification takes the form of dithered quantization.}. Consequently, much of the decoder’s functionality can be shifted to the encoder side. From this perspective, the problem reduces essentially to a channel synthesis (also known as channel simulation) task \cite{Cuff13}, for which the strong functional representation lemma \cite{LEG18} provides a convenient upper bound on $H(\hat{X}|K)$, which is often used as a proxy for $R$ in \eqref{eq:variable_length}.

\section{The Asymptotic Setting}\label{sec:asymptotic}

In contrast to the one-shot setting, which operates on a single source variable (albeit potentially high-dimensional), this section considers the asymptotic regime where a long sequence of i.i.d. source variables is processed jointly. As the blocklength grows, certain regularities begin to emerge, leading to notable simplifications---particularly when the dimension of each source variable is low---and enabling a more structured approach to system design. These patterns provide additional insight into the interplay between rate, distortion, and perception in the large-blocklength limit. To facilitate the discussion, we begin by introducing the relevant system definitions.

Let $\{X_t\}_{t=1}^{\infty}$ be an i.i.d. source process with marginal distribution $p_X$ over $\mathbb{R}^d$, and assume that $p_X$ is square-integrable, i.e., $\mathbb{E}[\|X\|^2]<\infty$. A length-$n$ perception-aware lossy source coding system consists of a stochastic encoder $f^{(n)}:\mathbb{R}^{d\times n}\times\mathcal{K}\rightarrow\mathcal{J}$, a stochastic decoder $g^{(n)}:\mathcal{J}\times\mathcal{K}\rightarrow\mathbb{R}^{d\times n}$, and a shared random seed $K$, which is uniformly distributed over $\mathcal{K}$ and independent of the source.
The  encoder $f^{(n)}$ maps  the source sequence $X^n$, together with the seed $K$, to an index $J$ in  $\mathcal{J}$ according to a conditional distribution $p_{J|X^nK}$, while the decoder $g^{(n)}$ produces the reconstruction $\hat{X}^n$ based on $(J,K)$ according to a conditional distribution $p_{\hat{X}^n|JK}$.

\subsection{Fundamental Limits}
	A distortion loss $D$ is said to be achievable under a compression rate constraint $R$, a common randomness rate constraint $C$, and a perception constraint $P$ 	
	if, for all sufficiently large blocklengths $n$, there exists a length-$n$ perception-aware lossy source coding system satisfying	
	\begin{align}
		&\frac{1}{n}\log|\mathcal{J}|\leq R,\label{eq:rate}\\
		&\frac{1}{n}\log|\mathcal{K}|\leq C,\label{eq:commonrandomness}\\
		&\frac{1}{n}\mathbb{E}[\|X^n-\hat{X}^n\|^2]\leq D,\label{eq:distortion}\\
		&\frac{1}{n}W^2_2(p_{X^n},p_{\hat{X}^n})\leq P.\label{eq:perception}
	\end{align}
	The infimum of all such achievable distortion losses is denoted by  $D(R,C,P)$.

As shown in \cite[Theorem 1]{QCYX25}, $D(R,C,P)$ admits the following information-theoretic characterization:
\begin{align}
	D(R,C,P)=&\min\limits_{p_{\tilde{X}}\in\mathcal{P}(\mathbb{R}^d), \mu,\nu\in\Pi(p_X,p_{\tilde{X}})}\mathbb{E}_{\mu}[\|X-\tilde{X}\|^2]+\left[\left(\sqrt{\mathbb{E}_{\nu}[\|X-\tilde{X}\|^2]}-\sqrt{P}\right)_+\right]^2\label{eq:inf}\\
	&\mbox{s.t.}\quad\mathbb{E}_{\mu}[X|\tilde{X}]=\tilde{X}\quad\mu\mbox{-a.s.},\label{eq:conditionexp}\\
	&\hspace{0.3in}I_{\mu}(X;\tilde{X})\leq R,\label{eq:constraintR}\\
	&\hspace{0.3in}I_{\nu}(X;\tilde{X})\leq R+C.\label{eq:constraintRc}	
\end{align}
Despite its conceptual elegance, the proof of this result is technically involved, relying heavily on tools such as soft-covering lemmas. As a consequence, the operational meaning of this characterization may not be immediately transparent, especially to readers without a strong background in information theory. In this tutorial, we aim to provide a more accessible perspective by elucidating the underlying coding scheme and relating it to conventional lossy source coding principles.

\subsection{Random Coding Scheme}

The following provides a sketch of the random coding argument underlying the characterization of 
$D(R,C,P)$. To avoid the technical difficulties associated with advanced tools such as soft-covering lemmas, we instead adopt a more transparent description based on concepts and constructions commonly used in classical rate–distortion theory. This approach allows us to convey the key ideas and intuition behind the coding scheme without delving into the full technical machinery required for a rigorous proof.
\begin{itemize}
\item Codebook generation: Let $\mu$ be a coupling in $\Pi(p_X,p_{\tilde{X}})$ satisfying
\begin{align}
&\mathbb{E}_{\mu}[X|\tilde{X}]=\tilde{X}\quad\mu\mbox{-a.s.},\label{eq:mummse}\\
&I_{\mu}(X;\tilde{X})\approx R.
\end{align}
Construct a codebook $\mathcal{C}:=\{\tilde{X}^n(j,k): j=0,1,\ldots,\lfloor 2^{nR}\rfloor-1, k=0,1,\ldots,\lfloor 2^{nC}\rfloor-1\}$, where  each codeword $\tilde{X}^n(j,k)$ is drawn independently according to the product distribution $p^n_{\tilde{X}}$. The codebook $\mathcal{C}$ is then partitioned into $2^{nC}$ subcodebooks $\mathcal{C}(k):=\{\tilde{X}^n(j,k):j=0,1,\ldots,\lfloor 2^{nR}\rfloor-1\}$, indexed by $k\in\{0,1,\ldots,\lfloor 2^{nC}\rfloor-1\}$. With high probability, each $\mathcal{C}(k)$ constitutes a good lossy source code for the coupling $\mu$.

% $2^{nC}$ codebooks, each consisting of $2^{nR}$ length-$n$ codewords

\item Encoding: Given the source sequence $X^n$ and the random seed $K$, the encoder selects an index $J$ that the corresponding codeword $\tilde{X}^n(J,K)$ in the
subcodebook $\mathcal{C}(K)$ is joint typicality with $X^n$ under $\mu$, or satisfies an appropriate proxy criterion.
 This encoding operation mirrors that of conventional lossy source coding. As a consequence, 
\begin{align}
\frac{1}{n}\mathbb{E}[\|X^n-\tilde{X}^n(J,K)\|^2]\approx\mathbb{E}_{\mu}[\|X-\tilde{X}\|^2].\label{eq:distortion_mu}
\end{align}

\item Decoding: Given $(J,K)$, the decoder first recovers the codeword $\tilde{X}^n(J,K)$. By the properties of good lossy source codes for the coupling $\mu$ \cite[Problem 10.9]{CT91} (see also \cite{ASP24}),
\begin{align} p_{X^n|\tilde{X}^n(J,K)}\approx\prod\limits_{t=1}^n\mu_{X|\tilde{X}},\label{eq:mufactorization}
\end{align}
which, together with \eqref{eq:mummse}, yields
\begin{align}
	\mathbb{E}[X^n|\tilde{X}^n(J,K)]\approx \tilde{X}^n(J,K).\label{eq:MMSE}
\end{align}
At this point, the procedure coincides with conventional lossy source coding; in the perception-aware framework, this corresponds to the combined operation of encoding followed by MMSE estimation.

Consider a transport plan that attains $W^2_2(p_{X^n},p_{\tilde{X}^n(J,K)})$. This  plan induces a lossy source coding scheme for $X^n$ based on the full codebook $\mathcal{C}$. Consequently, there exists a coupling\footnote{Roughly speaking, the coupling $\nu$ induced by a transport plan attaining $W^2_2(p_{X^n},p_{\tilde{X}^n(J,K)})$ corresponds to  one that minimizes $\mathbb{E}_{\nu}[\|X-\tilde{X}\|^2]$ subject to the constraint $I_{\nu}(X;\tilde{X})\leq R+C$. Other  couplings $\nu\in\Pi(p_X,p_{\tilde{X}})$ satisfying the same constraint are, in general, associated with suboptimal transport plans.
} $\nu\in\Pi(p_X,p_{\tilde{X}})$ such that
\begin{align}
	&I_{\nu}(X;\tilde{X})\approx R+C,\\
	&\mathbb{E}_{\nu}[\|X-\tilde{X}\|^2]\approx\frac{1}{n}W^2_2(p_{X^n},p_{\tilde{X}^n(J,K)}).\label{eq:distortion_nu}
\end{align}
We refer to the encoding associated with the full codebook $\mathcal{C}$ as virtual encoding, since it is not actually implemented. This should be contrasted\footnote{On the other hand, in light of Proposition~\ref{prop:1}, when common randomness is absent, there is no need to distinguish between virtual encoding and operational encoding, provided that the latter is optimized.} with the operational encoding, which is performed using the subcodebook $\mathcal{C}(K)$.

%first say optimal transport, then say it can be viewed as virtual encoding

%One can also think of $\tilde{X}^n(J,K)$ as the encoding output based on codebook $\mathcal{C}$. We shall refer to this encoding as virtual encoding as it is never performed. It should be contrasted with operational encoding based on subcodebook $\mathcal{C}(K)$.

To clarify the distinction between virtual encoding and operational encoding, it is instructive to revisit the unit-circle example, in particular the case of 
$1$-bit quantization with 
$1$-bit of common randomness (see also Fig. \ref{fig:operational_vs_virtual}). In this setting, the centroids of the right and left unit semicircles form one subcodebook, while those of the upper and lower semicircles form another; together, these four centroids constitute the full codebook 
$\mathcal{C}$. Operational encoding uses only one subcodebook at a time, as determined by the realization of the shared random seed, and never the entire codebook jointly. By contrast, if the full codebook 
$\mathcal{C}$ were used for encoding, the unit circle would be partitioned into four quarter circles, each mapped to its nearest centroid. Although this virtual encoding is not implemented in practice, it plays a crucial role in guiding the decoder design. In particular, to generate a perceptually perfect sample, it suffices to map each centroid back to a uniform distribution over its corresponding quarter circle, which coincides with the stochastic inverse of the virtual encoding.

Returning to the general case, 
we first generate from $\tilde{X}^n(J,K)$
a perceptually perfect sample $X'^n$ according to the conditional distribution $q_{X^n|\tilde{X}^n(J,K)}$ induced by the aforementioned transport plan attaining $W^2_2(p_{X^n},p_{\tilde{X}^n(J,K)})$. Notably, $q_{X^n|\tilde{X}^n(J,K)}$ can be viewed as the stochastic inverse of the virtual encoding based on the full codebook $\mathcal{C}$. In analogy with \eqref{eq:mufactorization},
we have
\begin{align}
	q_{X^n|\tilde{X}^n(J,K)}\approx\prod\limits_{t=1}^n\nu_{X|\tilde{X}},\label{eq:factorization}
	\end{align}
which implies that $X'^n$  can be generated from $\tilde{X}^n(J,K)$ 
approximately in a symbol-by-symbol manner.
Linearly interpolating $\tilde{X}^n(J,K)$ and $X'^n$ yields the reconstruction
\begin{align}
	\hat{X}^n:=(1-\alpha)X'^n+\alpha\tilde{X}^n(J,K),
\end{align}
where
\begin{align}
	\alpha:=\begin{cases}
		\frac{\sqrt{nP}}{W_2(p_{X^n},p_{\tilde{X}^n(J,K)})}, & nP\in[0,W^2_2(p_{X^n},p_{\tilde{X}^n(J,K)})),\\
		1, & nP\geq W^2_2(p_{X^n},p_{\tilde{X}^n(J,K)}).
	\end{cases}
	\end{align}
The resulting end-to-end distortion satisfies
\begin{align}
	\frac{1}{n}\mathbb{E}[\|X^n-\hat{X}^n\|^2]&\stackrel{(a)}{\approx}\frac{1}{n}\mathbb{E}[\|X^n-\tilde{X}^n(J,K)\|^2]+\left[\left(\frac{W_2(p_{X^n},p_{\tilde{X}(J,K)})}{\sqrt{n}}-\sqrt{P}\right)_+\right]^2\nonumber\\
	\nonumber\\
	&\stackrel{(b)}{\approx}\mathbb{E}_{\mu}[\|X-\tilde{X}\|^2]+\left[\left(\sqrt{\mathbb{E}_{\nu}[\|X-\tilde{X}\|^2]}-\sqrt{P}\right)_+\right]^2,
\end{align}
where  ($a$) follows from \eqref{eq:MMSE} and Proposition \ref{prop:interpolation}, and ($b$) from \eqref{eq:distortion_mu} and \eqref{eq:distortion_nu}. Finally, it can be verified by invoking the same line of argument as  in \eqref{eq:perception1} and \eqref{eq:perception2}
that the perception constraint $\frac{1}{n}W^2_2(p_{X^n},p_{\hat{X}^n})\leq P$ is satisfied. This portion of the random coding scheme corresponds to the generator and interpolator modules in the perception-aware lossy source coding system.
\end{itemize}

%Returning to the general case, what relevant is the backward channel, denoted by $\nu_{X^n|\tilde{X}}$, induced by virtual encoding. 

%comment on $\mathbb{E}[\|X'^n-\tilde{X}^n(J,K)\|^2]$
%vs. $W^2_2(p_{X^n},p_{\tilde{X}^n(J,K)})$

%comment on factorization

%We can use $\nu_{X^n|\tilde{X}}$ generate a perceptually perfect sample $X'^n$ based on $\tilde{X}^n(J,K)$. Then construct reconstruction $X^n$ by linearly interpolating $X'^n$ and $\tilde{X}^n(J,K)$.

Alert readers may have noticed the close resemblance between the above scheme and its one-shot counterpart. This similarity is not surprising, as a long sequence of source variables can be viewed as a single ``super" source variable, to which the one-shot scheme can be directly applied. At the same time, the asymptotic setting brings additional structure that clarifies the connection between perception-aware and conventional lossy source coding. In particular, it highlights the role of the backward channel $q_{X^n|\tilde{X}^n(J,K)}$ in generating perceptually perfect samples, and reveals the simplifications that arise from its factorization \eqref{eq:factorization}, thereby enabling a more structured and interpretable design.

\subsection{Gaussian Case}

For the scalar Gaussian case $X\sim\mathcal{N}(0,1)$, the optimization problem in \eqref{eq:inf}--\eqref{eq:constraintRc} can be solved explicitly, yielding \cite[Theorem 2]{QCYX25}
\begin{align}
	D(R,C,P)=2^{-2R}+\left[\left(\sqrt{2-2^{-2R}-2\sqrt{(1-2^{-2R})(1-2^{-2(R+C)})}}-\sqrt{P}\right)_+\right]^2.\label{eq:Gaussian}
\end{align}
This result admits a geometric interpretation analogous to that of the unit-circle example.
Specifically, let $\mathbb{S}$ and $\tilde{\mathbb{S}}$ be $(n-1)$-dimensional spheres centered at the origin with radii $\sqrt{n}$ and $\sqrt{n(1-2^{-2R})}$, respectively. Owing to the concentration (hardening) phenomenon, $X^n$ is approximately uniformly distributed over the surface of  $\mathbb{S}$.
For each $j\in\{0,1,\ldots,\lfloor 2^{nR}\rfloor-1\}$ and $k\in\{0,1,\ldots,\lfloor 2^{nC}\rfloor-1\}$, draw $\tilde{X}^n(j,k)$ independently and uniformly from the surface of $\tilde{\mathbb{S}}$.
Let $x^n$ be an arbitrary point on the surface of $\mathbb{S}$. We have
\begin{align}
	\|x^n-\tilde{X}^n(j,k)\|^2&=\|x^n\|^2+\|\tilde{X}^n(j,k)\|^2-2\langle x^n,\tilde{X}^n(j,k)\rangle\\
	&=n+n(1-2^{-2R})-2n\sqrt{1-2^{-2R}}\frac{\langle x^n,\tilde{X}^n(j,k)\rangle}{\|x^n\|\|\tilde{X}^n(j,k)\|}.
\end{align}
Standard concentration arguments imply that, with high probability,
\begin{align}
	&\min\limits_{j\in\{0,1,\ldots,\lfloor 2^{nR}\rfloor-1\}}\frac{\langle x^n,\tilde{X}^n(j,k)\rangle}{\|x^n\|\|\tilde{X}^n(j,k)\|}\approx\sqrt{1-2^{-2R}},\quad k=0,1,\ldots,\lfloor 2^{nC}\rfloor-1,\\
	&\min\limits_{j\in[2^{nR}],k\in\{0,1,\ldots,\lfloor 2^{nC}\rfloor-1\}}\frac{\langle x^n,\tilde{X}^n(j,k)\rangle}{\|x^n\|\|\tilde{X}^n(j,k)\|}\approx\sqrt{1-2^{-2(R+C)}}.
\end{align}
Therefore, with high probability,
\begin{align}
	&\min\limits_{j\in\{0,1,\ldots,\lfloor 2^{nR}\rfloor-1\}}	\frac{1}{n}\|x^n-\tilde{X}^n(j,k)\|^2\approx 2^{-2R},\quad k=0,1,\ldots,\lfloor 2^{nC}\rfloor-1,\label{eq:operational}\\
	&\min\limits_{j\in\{0,1,\ldots,\lfloor 2^{nR}\rfloor-1\},k\in\{0,1,\ldots,\lfloor 2^{nC}\rfloor-1\}}	\frac{1}{n}\|x^n-\tilde{X}^n(j,k)\|^2\approx 2-2^{-2R}-2\sqrt{(1-2^{-2R})(1-2^{-2(R+C)})},\label{eq:virtual}
\end{align}
which correspond to the mean squared errors of operational encoding and virtual encoding, respectively. Substituting these two mean squared errors into the distortion expression of the interpolation scheme gives
\eqref{eq:Gaussian}.

\subsection{Structured Code Construction}

While random coding arguments are instrumental in establishing fundamental information-theoretic limits, they are typically impractical for implementation. In practice, one can instead employ structured codes to realize the same architectural principles in a more efficient and scalable manner. This replacement preserves the essential behavior dictated by the underlying theory, while enabling more scalable practical implementations.

To this end, assume $p_{\tilde{X}}$ is supported over a finite set, which incurs no essential loss of generality under appropriate discretization. Through relabeling and potential zero-padding, we can map  $p_{\tilde{X}}$ to a distribution $p_{U}$ defined over the cyclic group $\mathbb{Z}_q$ for some prime $q$. Consequently, the lossy source coding problem associated with a coupling $\mu\in\Pi(p_X,p_{\tilde{X}})$
 is transformed into an equivalent problem involving a coupling $\mu'\in\Pi(p_X,p_{U})$. This reformulated problem can then be implemented using nested Abelian group
  codes \cite{KP11}. The new ingredient here is the need to employ different lossy source codebooks for different realizations of the random seed 
$K$ in order to leverage the benefits of common randomness, while ensuring uniform performance across all such realizations. To address this, consider a chain of three nested Abelian group codes, $\mathcal{C}_1 \subseteq \mathcal{C}_2 \subseteq \mathcal{C}_3$. Their respective rates are set to approximately $\log q - H(U)$, $\log q - H(U) + R$, and $\log q - H(U) + R + C$, with $R \approx I(X; U)$. We select a suitable system of coset representatives for the partition of the ambient space $\mathbb{Z}^n_q$ induced by 
$\mathcal{C}_1$
such that the resulting representative set almost completely covers the typical set of 
$p_U$.
Within this set, the codewords of $\mathcal{C}_3$ form approximately $2^{R+C}$ coset representatives of $\mathcal{C}_1$ in $\mathcal{C}_3$. 
These representatives are then partitioned into $2^C$ distinct classes based on their membership in the cosets of $\mathcal{C}_2$ relative to $\mathcal{C}_1$. 
In this construction, each class constitutes a sub-codebook $\mathcal{C}(k)$, while their union defines the full codebook $\mathcal{C}$.

\begin{figure}[htbp]
	\centerline{\includegraphics[width=15cm]{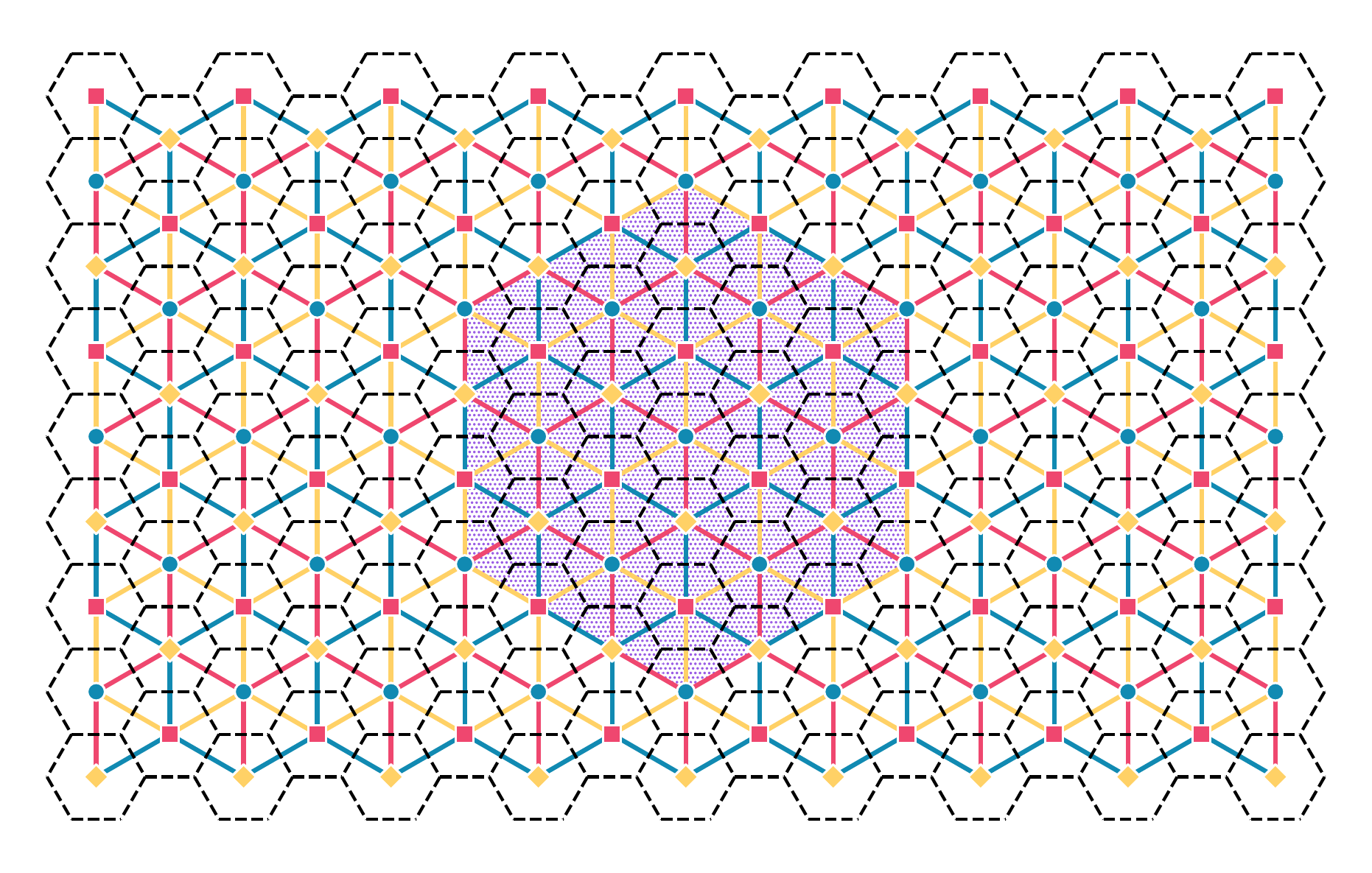}} \caption{
		Illustration of a lattice-based construction. The blue points represent the intermediate lattice $\Lambda_2$, while the yellow and red points denote its cosets; together, they form the finest lattice $\Lambda_3$. The Voronoi regions of points in $\Lambda_2$
 (or its cosets) are delineated by solid lines in the corresponding colors and are associated with operational encoding,  which essentially amounts to conventional lattice quantization, mapping each region to its lattice point; the choice of coset depends on the realization of the random seed $K$. The Voronoi regions of points in the finest lattice $\Lambda_3$, shown with dashed lines, are associated with virtual encoding, whose stochastic inverse is used to generate perceptually perfect samples. The shaded part represents the fundamental Voronoi region $\mathcal{V}_1$ of the shaping lattice $\Lambda_1$. Bit representations are restricted to points in $\Lambda_2$ (or its cosets) that lie within $\mathcal{V}_1$.}\label{fig:lattice}
\end{figure}

The nested Abelian group code hierarchy can be translated into a lattice-based construction (see Fig. \ref{fig:lattice}) by transitioning from a discrete finite group to a continuous Euclidean space. In this geometric framework, the nested codes $\mathcal{C}_1 \subseteq \mathcal{C}_2 \subseteq \mathcal{C}_3$ are replaced by a chain of nested lattices $\Lambda_1 \subseteq \Lambda_2 \subseteq \Lambda_3$. The role of the coarsest code $\mathcal{C}_1$ is assumed by the shaping lattice $\Lambda_1$, whose fundamental Voronoi region $\mathcal{V}_1$, under appropriate scaling, is designed to capture the high-probability typical set of the source distribution. Bit representations are restricted to the points of the intermediate lattice $\Lambda_2$ (or its cosets in $\Lambda_3$) that lie within $\mathcal{V}_1$. This restriction is necessitated by the fixed-length coding assumption; if variable-length (entropy) coding is employed, the boundary defined by $\Lambda_1$ becomes unnecessary\footnote{A similar observation applies to the coarsest code $\mathcal{C}_1$ in the nested Abelian group construction.}. The intermediate lattice 
$\Lambda_2$ determines the quantization
structure: each of its cosets in 
$\Lambda_3$ serves as a candidate quantizer, with the choice governed by the realization of the random seed 
$K$. The union of these cosets forms the finest lattice 
$\Lambda_3$, which leverages the available common randomness to improve
 the distortion–perception tradeoff.
 Just as in the unit-circle example, when unlimited common randomness is available (i.e., 
$C\rightarrow\infty$), the construction reduces to dithered quantization, as the finest lattice effectively becomes the ambient Euclidean space.

It should be emphasized that the above description is intended only to provide high-level architectural guidelines. The detailed implementation of structured codes for perception-aware lossy source coding, along with a rigorous performance analysis, remains a challenging and worthwhile endeavor.
See \cite{LHS25} for a recent attempt in this direction.

%The above nested linear code construction can be translated to a lattice construction

 %by employing a chain of nested lattices $\Lambda_1 \subseteq \Lambda_2 \subseteq \Lambda_3$.

%virtual coding becomes operational coding when no common randomness is available.

%corresponds to choosing the optimal coupling $\nu$

%some structural patterns emerge, no training (esp. when $d$ is small)

%structured codes can be translated to lattice constructions. See [] for some recent attempt in this direction.

 %which corresponds to the distortion loss and the perception loss, respectively. Substituting these two losses into the interpolation formula and normalizing by $n$ gives
%\begin{align*}
%	D(R,C,P)=2^{-2R}+\left[\left(\sqrt{2-2^{-2R}-2\sqrt{(1-2^{-2R})(1-2^{-2(R+C)})}}-\sqrt{P}\right)_+\right]^2,
%\end{align*}
%which coincides with the distortion-rate-perception function with limited common randomness for the quadratic Wasserstein Gausssian case. This is not surprising, as high-dimensional Gaussian vectors concentrate sharply near the surface of a sphere.

\subsection{Asymptotic Universal Representation}

The notion of universal representation can be extended from the one-shot setting to the asymptotic setting. We say an asymptotic universal representation exists if the minimizer $(p_{\tilde{X}},\mu,\nu)$ of the optimization problem in \eqref{eq:inf}--\eqref{eq:constraintRc}  does not depend on the perception level $P$. This property holds trivially in the absence of common randomness (i.e, $C=0$), since the asymptotic notion is weaker than its one-shot counterpart. More interestingly, it has been shown \cite{QCYX25} that, in the scalar Gaussian case, an asymptotic universal representation exits  even when common randomness is available (i.e., $C>0$).  In contrast, for the vector Gaussian case with common randomness, the existence of such a representation is no longer guaranteed; nevertheless, the penalty incurred by adopting a perception-oblivious encoder, as characterized by the classical reverse-waterfilling formula, appears to be rather small.

%asymptotic universal representation

%suboptimality of posterior sampling

\section{Discussion}\label{sec:conclusion}

In this tutorial, we have distilled several key guidelines for the design of perception-aware lossy source coding systems. These guidelines provide a structured foundation for integrating deep learning techniques within a principled framework, rather than treating the entire system as a black box and relying solely on blind end-to-end training. By elucidating the roles of different components and their interplay, this perspective offers actionable insights for systematic design. It is our hope that this tutorial will foster the development of constructive approaches with tangible real-world impact on learned image compression.

%We have mainly focused on the squared-error distortion measure and the squared Wasserstein-$2$ perception measure. It is worth mentioning that with a minor adjustment of the interpolation parameter (see Fig.), the current framework can be readily applied to other perception measures, although no optimality is guaranteed anymore (but it is reasonable to expect that this framework continues to work reasonably well as long as the geometry associated with the adopted perception measure is not too different from that under the Wasserstein-$2$ distance). On the other hand, extension to other distortion measures likely require more substantial changes. Overall, by pinpointing the limitation of the current framework, we hope to trigger further research on a more general characterization of the rate-distortion-perception limits.

\begin{figure}[htbp]
	\centerline{\includegraphics[width=12cm]{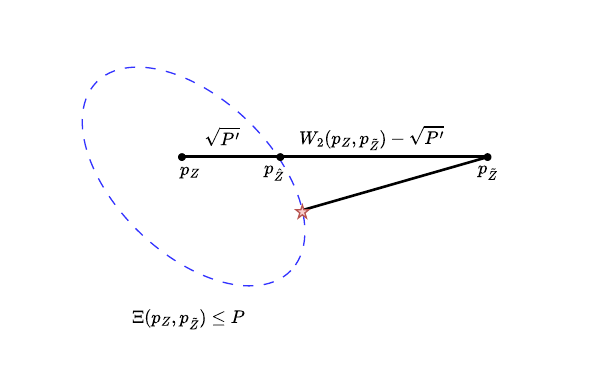}} \caption{Consider a random vector $Z$ in $\mathbb{R}^k$ with $\mathbb{E}[\|Z\|^2]<\infty$, and a representation $W$ generated according to some conditional distribution $p_{W|Z}$.
		Let $\tilde{Z}$ be the MMSE estimate of  $Z$ based on  $W$, and let $\hat{Z}$ be a reconstruction derived from  $W$.
		For a given perception measure
		$\Xi$, the constraint $\Xi(p_Z,p_{\hat{Z}})\leq P$ defines a perceptually permissible set (illustrated as an elliptical region) within which
		$p_{\hat{Z}}$ must lie. The interpolation scheme selects
		$p_{\hat{Z}}$ as the point where the Wasserstein-$2$ geodesic  from $p_Z$ to $p_{\tilde{Z}}$ intersects the boundary of this set. For this choice,  $W_2(p_Z,p_{\hat{Z}})=\sqrt{P'}$, where $P'$ may differ from $P$ when  $\Xi$ is not the squared Wasserstein-$2$ distance. However, this choice is generally not optimal in minimizing $W_2(p_{\hat{Z}},p_{\tilde{Z}})$ over the perceptually permissible set (as indicated by the starred point). By instead generating $\hat{Z}$ based on $\tilde{Z}$ via a  Wasserstein-$2$ optimal transport plan that maps $p_{\tilde{Z}}$ to this optimal point, one obtains a smaller value of
		$W_2(p_{\hat{Z}},p_{\tilde{Z}})$, and consequently a lower end-to-end distortion $\mathbb{E}[\|Z-\hat{Z}\|^2]$.
		See Fig. \ref{fig:interpolation} for a comparison.}\label{fig:interpolation_2}
\end{figure}

We have primarily focused on the squared-error distortion measure and the squared Wasserstein-$2$ perception measure. It is worth noting that, with a minor adjustment of the interpolation parameter, the proposed framework can be extended to accommodate other perception measures, although optimality is no longer guaranteed (see Fig. \ref{fig:interpolation_2}). Nevertheless, it is reasonable to expect that the framework remains effective as long as the underlying geometry induced by the chosen perception measure does not deviate significantly from that of the Wasserstein-$2$ distance. In contrast, extending the framework to other distortion measures is likely to require more substantial modifications. By clearly delineating these limitations, we hope to stimulate further research toward a more general characterization of rate–distortion–perception tradeoffs, along with the associated architectural principles.

We conclude this tutorial by highlighting some promising directions for future research. First, among all components of the current framework, the generator module for realizing a generic Wasserstein-$2$ optimal transport remains the least satisfactorily implemented. While recent advances such as rectified flow represent an important step forward, substantial room for improvement remains, and further progress in this area is likely to have implications well beyond perception-aware lossy source coding. Second, perception-aware lossy source coding in the asymptotic setting calls for integrating structured codes into modern learning-based systems or, conversely, incorporating learning-based components into coding architectures to enable joint design and optimization. Such a hybrid approach has the potential to combine the theoretical guarantees and interpretability of structured coding schemes with the flexibility and data-driven adaptability of machine learning models. Developing principled frameworks for this integration, together with efficient training and implementation strategies, could lead to significant improvements in performance, scalability, and robustness across a wide range of applications.
Last but not least, the formulation of the perception measure itself warrants re-examination \cite{QWBT24,Chen26}. To date, most information-theoretic work on rate-distortion-perception theory has focused on adopting various divergences as proxies for perceptual quality; however, the operational meaning of such distribution-based measures is not yet fully understood. In particular, they do not readily support the assessment of perceptual quality at the level of individual samples. It also remains unclear whether the current design framework, developed under distribution-based perception measures, retains its robustness when evaluated using alternative perception criteria.

%common randomness 0, double distortion

%common randomness $\infty$, connecting two points

\end{document}